\documentclass[sageh,times,doublespace]{sagej}
\usepackage{moreverb,url}
\usepackage[colorlinks,bookmarksopen,bookmarksnumbered,citecolor=red,urlcolor=red]{hyperref}
\newcommand\BibTeX{{\rmfamily B\kern-.05em \textsc{i\kern-.025em b}\kern-.08em
\kern-.1667em\lower.7ex\hbox{E}\kern-.125emX}}
\def\volumeyear{2023}
\usepackage[font=bf,skip=\baselineskip]{caption} 
\usepackage{multirow}
\usepackage{xspace} 
\usepackage{booktabs}
\usepackage{seqsplit}
\usepackage{adjustbox}
\usepackage{graphicx}

\usepackage[shortcuts]{glossaries}
\glsdisablehyper
\newacronym{bih}{BIH}{bounding interval hierarchy}
\newacronym[shortplural={BVHs},longplural={bounding volume hierarchies}]{bvh}{BVH}{bounding volume hierarchy}
\newacronym{csg}{CSG}{constructive solid geometry}
\newacronym{gcd}{GCD}{graphics complex die}
\newacronym{hep}{HEP}{high energy physics}
\newacronym{hpc}{HPC}{high performance computing}
\newacronym{mc}{MC}{Monte Carlo}
\newacronym{orange}{ORANGE}{Oak Ridge Adaptable Nested Geometry Engine}
\newacronym{ornl}{ORNL}{Oak Ridge National Laboratory}
\newacronym{rtk}{RTK}{Reactor Tool Kit}
\newacronym{sah}{SAH}{surface area heuristic}
\newacronym{simt}{SIMT}{single instruction, multiple threads}
\newacronym{smr}{SMR}{small modular reactor}
\newacronym{pwr}{PWR}{pressurized water reactor}

\newcommand{\keff}{\ensuremath{k_{\mathrm{eff}}}}
\newcommand{\fsrc}{\ensuremath{\text{f}}}
\newcommand{\op}[1]{\hat{\mathrm{#1}}}
\newcommand{\findcell}{find\_cell\xspace}
\newcommand{\crosssurface}{cross\_surface\xspace}
\newcommand{\distancetosurface}{distance\_to\_surface\xspace}
\newcommand{\changedirection}{change\_direction\xspace}
\newcommand{\movewithincell}{move\_within\_cell\xspace}
\newcommand{\gettracker}{get\_tracker\xspace}
\newcommand{\getrecttracker}{get\_rect\_tracker\xspace}
\newcommand{\getcsgtracker}{get\_csg\_tracker\xspace}
\newcommand{\runhistories}{run\_histories\xspace}

\newcommand{\visittracker}{visit\_tracker\xspace}
\newcommand{\cpp}{C\texttt{++}\xspace}

\usepackage{algorithm}
\usepackage[noend]{algpseudocode}
\algrenewcommand\algorithmicprocedure{{\bf procedure}}
\algrenewcommand\algorithmicfor{{\bf for}}
\algrenewcommand\algorithmicif{{\bf if}}
\algrenewcommand\algorithmicelse{{\bf else}}
\algrenewcommand\algorithmicend{{\bf end}}
\algrenewcommand\algorithmicreturn{{\bf return}}
\algnewcommand\Not{{\bf not }}
\algrenewcommand{\algorithmiccomment}[1]{{$//$ #1}}

\algrenewcommand\alglinenumber[1]{\footnotesize #1}

\usepackage{listings}
\renewcommand{\lstlistingname}{\bfseries Listing}
\makeatletter
\def\fnum@lstlisting{%
  \lstlistingname
  \ifx\lst@@caption\@empty\else~\thelstlisting\normalfont\fi}%
\makeatother

\usepackage{colortbl} 
\usepackage{xcolor}   
\definecolor{colcolor}{RGB}{255,239,228}

\begin{document}
\runninghead{Biondo et al.}
\title{Comparison of nested geometry treatments within GPU-based Monte Carlo
       neutron transport simulations of fission reactors
\footnote{This manuscript has been authored by UT-Battelle, LLC, under contract
DE-AC05-00OR22725 with the US Department of Energy. The United States
Government retains and the publisher, by accepting the article for
publication, acknowledges that the United States Government retains a
nonexclusive, paid-up, irrevocable, worldwide license to publish or
reproduce the published form of this manuscript, or allow others to do so,
for United States Government purposes. DOE will provide access to these
results of federally sponsored research in accordance with the DOE Public
Access Plan (http://energy.gov/downloads/doe-public-access-plan).} 
}

\author{Elliott Biondo, Thomas Evans, Seth Johnson, and Steven Hamilton}
\affiliation{Oak Ridge National Laboratory, Oak Ridge, TN, USA}
\corrauth{Elliott Biondo, Oak Ridge National Laboratory,
1 Bethel Valley Rd.,
Oak Ridge, TN, 37830, USA.}
\email{veb@ornl.gov}

\begin{abstract}
Monte Carlo (MC) neutron transport provides detailed estimates of radiological
quantities within fission reactors. This involves tracking individual
neutrons through a computational geometry. CPU-based MC codes use multiple
polymorphic tracker types with different tracking algorithms to exploit the
repeated configurations of reactors, but virtual function calls have high
overhead on the GPU. The Shift MC code was modified to support GPU-based
tracking with three strategies: dynamic polymorphism with virtual functions,
static polymorphism, and a single tracker type with tree-based acceleration.
On the Frontier supercomputer these methods achieve 77.8\%, 91.2\%, and 83.4\%,
respectively, of the tracking rate obtained using a specialized tracker
optimized for rectilinear-grid-based reactors. This indicates that all three
methods are suitable for typical reactor problems in which tracking does not
dominate runtime. The flexibility of the single tracker method is highlighted
with a hexagonal-grid microreactor problem, performed without
hexagonal-grid-specific tracking routines, providing a 2.19$\times$ speedup
over CPU execution.
\end{abstract}

\maketitle

\section{Introduction}\label{sec:introduction}

\begin{figure}
\centerline{\includegraphics[width=0.6\columnwidth]{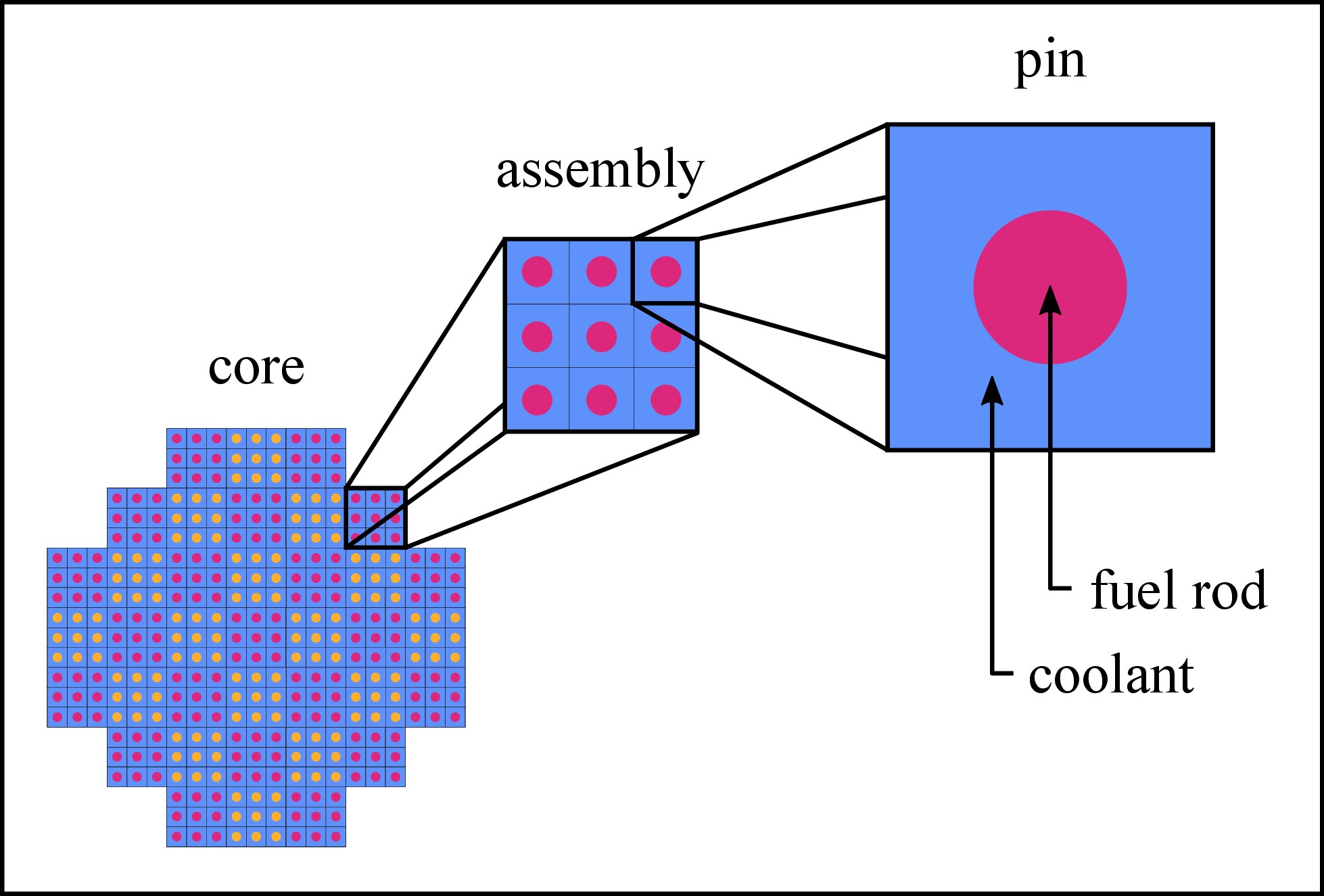}}
\caption{Simplified diagram of a generic \ac{pwr} core shown as a 2D slice.
Fuel rods are shown in red and yellow, ostensibly with different fuel
compositions.}
\label{fig:core}
\end{figure}

Nuclear reactors account for nearly 20\% of electricity production in the
United States, with lifecycle greenhouse gas emissions 17--29$\times$ less than
coal-fired power plants per unit of energy generated \citep{nuclear_facts}.
These reactors derive energy from induced nuclear fission, a process in which a
free neutron is captured by a heavy \emph{fuel} nucleus, causing it to split
into multiple lighter nuclei. Each fission releases heat, as well as additional
free neutrons that may induce subsequent fissions, perpetuating the process.
The \ac{pwr}---the most common reactor design worldwide---consists of a
\emph{core} composed of \emph{assemblies}, each of which is composed of fuel
\emph{rods}, as shown in Figure~\ref{fig:core}. Water is pumped through the
core to extract heat, and electricity is generated with a thermodynamic power
cycle via the expansion of steam through a turbine.

Significant computational modeling and simulation is required for reactor
design, licensing, and operation.  \ac{mc} neutron transport is the preeminent
method for obtaining high-fidelity estimates of radiological quantities because
of its continuous (i.e., non-discrete) treatment of space, direction, and
energy dimensions.  This stochastic method involves simulating neutron
\emph{histories}---the circuitous paths that individual neutrons take within a
reactor---using a random walk technique. By simulating a large number of
histories, accurate statistical estimates of radiological quantities can be
deduced. A key radiological quantity is the effective neutron multiplication
factor ($\keff$), defined as the average number of neutrons born from fission
that induce subsequent fissions.  A $\keff$ of 1 indicates that the reactor is
\emph{critical}, i.e., operating at steady state, whereas $\keff<1$ and $\keff >
1$ indicate that the fission rate will decrease or increase over time,
respectively. Estimates of $\keff$ are obtained using \ac{mc} via a power
iteration scheme \citep{lieberoth_power_1968}.

\begin{figure}
\centerline{\includegraphics[width=0.6\columnwidth]{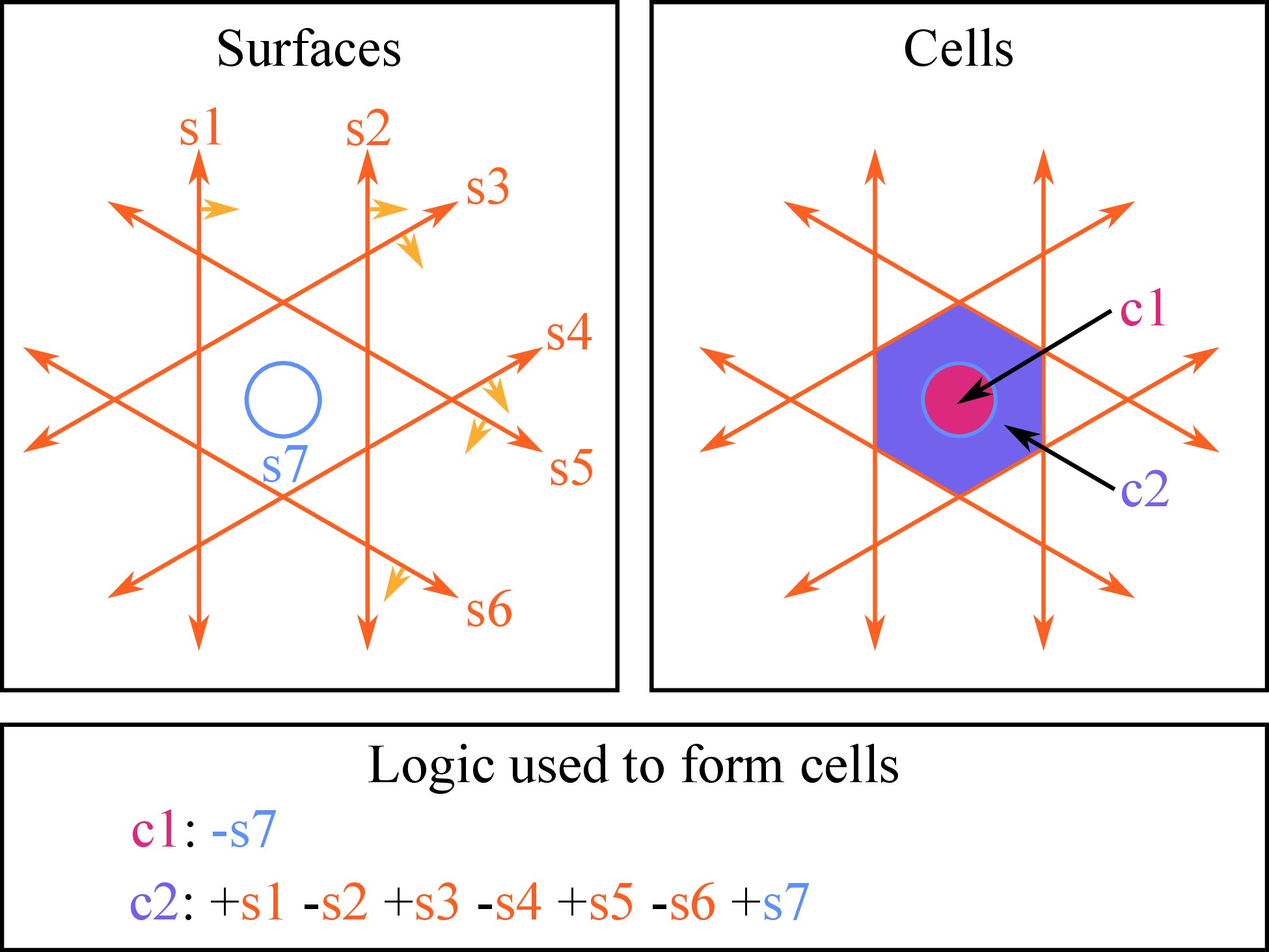}}
\caption{Demonstration of the \ac{csg} construction process, shown as a 2D
slice. Plane surfaces (orange) have surface normals (yellow) denoting which
side of the plane is positive. The outside of the blue cylindrical surface is
considered positive.  Cell construction logic is used to form cells 1 and 2.
The cell 2 logic can be read as ``the intersection of the space on the positive
side of surface 1, the negative side of surface 2, the positive side of surface
3, ... etc.''}
\label{fig:csg}
\end{figure}

Simulating neutron histories requires \emph{tracking} the positions of neutrons
within a computational representation of the reactor, usually a \ac{csg} model.
This is accomplished in a fashion similar to the ray tracing techniques used in
computer graphics rendering. With \ac{csg}, models are constructed from surface
primitives (e.g., planes, cylindrical shells, spherical shells) and Boolean
logic operations to form \emph{cells}, i.e., closed regions with uniform
material properties, as demonstrated in Figure~\ref{fig:csg}. Tracking
operations such as those summarized in Table~\ref{table:tracking} are
implemented by querying the surfaces that comprise each cell.

\begin{table*}
\small\sf\centering
\caption{High-level geometry tracking operations required by \ac{mc} transport codes.}
\label{table:tracking}
\begin{tabular}{lll}
\toprule
Tracking operation & Inputs               & Output \\
\midrule
\findcell          & (1) position          & (1) cell containing the position \\
\midrule
\distancetosurface & (1) position          & (1) distance to the next surface \\
                   & (2) direction         & (2) next surface \\
\midrule
\movewithincell    & (1) position          & (1) new position \\
                   & (2) direction         & \\
                   & (3) distance          & \\
\midrule
\crosssurface      & (1) position          & (1) new cell after crossing the surface \\
                   & (2) current cell      & \\
                   & (3) current surface   & \\
\midrule
\changedirection   & (1) new direction     & - \\    
\bottomrule
\end{tabular}
\end{table*}

\begin{figure}
\centerline{\includegraphics[width=0.6\columnwidth]{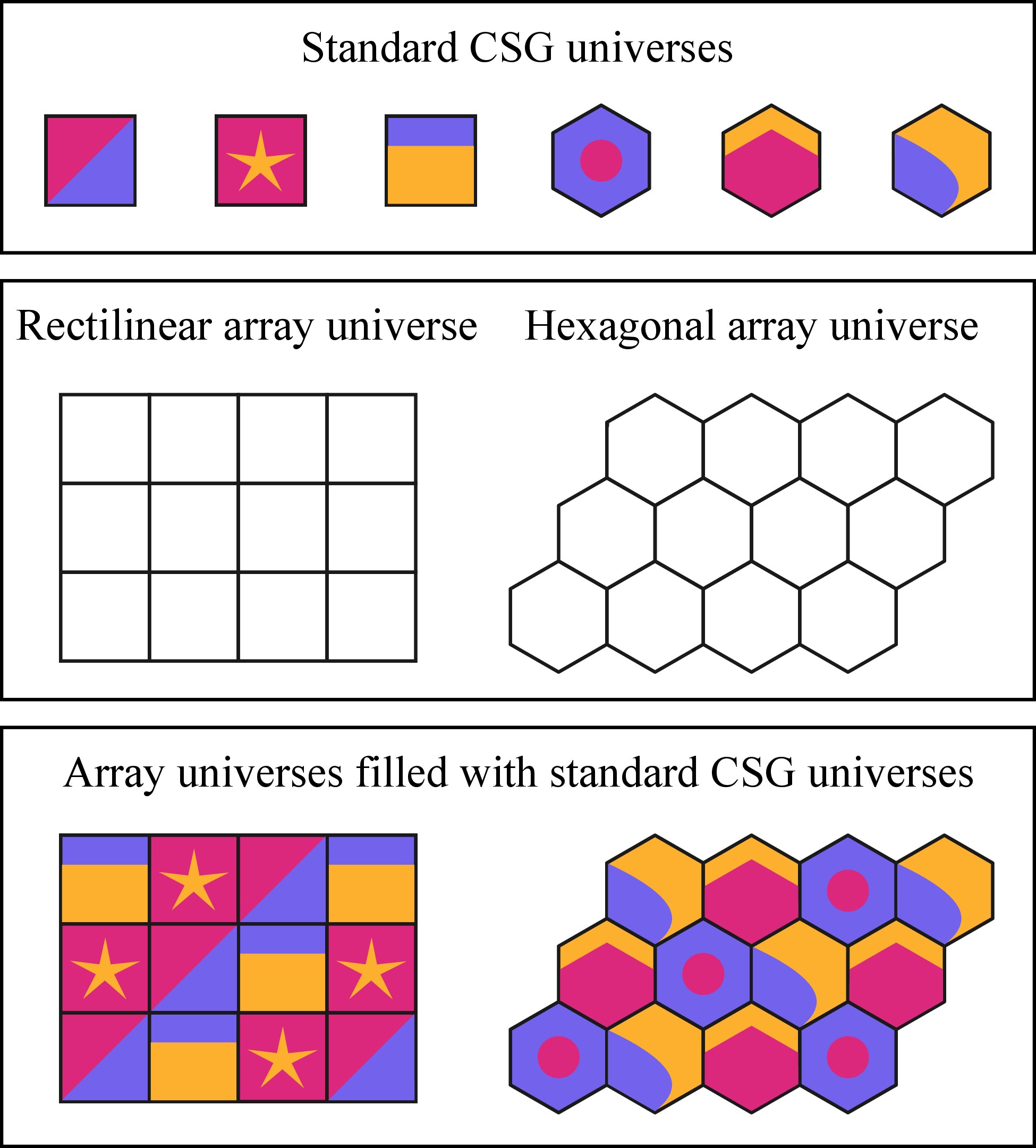}}
\caption{Demonstration of \ac{csg} universes nested within rectilinear and
hexagonal array universes, shown as a 2D slice.
The rectilinear array universe is shown with uniform spacing for simplicity, but
non-uniform spacing is required to support typical reactor applications.
}
\label{fig:universes}
\end{figure}

The nested and repeated structures that comprise reactors receive special
treatment for performance and user convenience. A \ac{csg} \emph{universe} is a
contiguous geometric region composed of one or more cells that can be embedded
within one or more \emph{parent} cells \citep{west_keno_1979}.  \emph{Array}
universes are special universe types in which cells comprise a structured mesh.
The cells in an array universe must be filled with other universes which may be
either \ac{csg} universes or other array universes. Examples of \ac{csg}
universes embedded in rectilinear and hexagonal array universes are shown in
Figure~\ref{fig:universes}.  Some fission reactors, including \acp{pwr}, can be
modeled with three \emph{levels} of universes: an array universe representing
the core, with each core array cell filled with an assembly array universe, and
each assembly array cell filled with a \ac{csg} universe containing fuel rods
surrounded by coolant (referred to as a \emph{pin} within this work).

The use of array universes in concert with \ac{csg} universes provides tracking
performance benefits. When determining what cell contains a given point, the
hierarchical configuration of nested universes can be exploited by first
finding which assembly contains the point, then which pin within the assembly,
and finally which cell within the pin.  This is loosely analogous to standard
ray tracing acceleration structures such as \acp{bvh} \citep{ericson_bvh_2004}
or $k$-d trees \citep{bentley_kd_1975}.  In addition, because \ac{csg} and
array universes represent geometry differently, the tracking operations
summarized in Table~\ref{table:tracking} are implemented separately for each
universe type, allowing each to be optimized.  Within this work, objects that
carry out these tracking operations are referred to as \emph{trackers}.  Thus,
a \ac{csg} universe has a corresponding \ac{csg} tracker, a rectilinear array
universe has a rectilinear array tracker, etc.  Array trackers can implement
tracking operations significantly more efficiently than \ac{csg} trackers by
exploiting the regular structure of array universes.

When using multiple tracker types, polymorphism can be employed to call tracking functions.
On the CPU with \cpp, this is accomplished by implementing
tracker functions as virtual functions of a tracker base class. However, using
GPU programming models (e.g., HIP, CUDA), it is not clear whether the
performance benefits of different tracker types outweigh the overhead
associated with virtual function calls \citep{zhang_virtual_2023}.

\emph{Shift} is a general-purpose \ac{mc} radiation transport code developed at
\ac{ornl} capable of simulating the behavior of neutrons and high-energy
photons for fission, fusion, and national security applications
\citep{pandya_shift_2016}. Shift, written in \cpp with abstracted HIP/CUDA
device programming models, supports both CPU and GPU execution
\citep{hamilton_gpu_2019} and is designed to scale effectively from laptops to
leadership-class supercomputers. On the CPU, Shift uses the \ac{orange}
\citep{johnson_orange_2023},\footnote{The \ac{orange} package is shared between
Shift and the Celeritas \ac{mc} code, used for high-energy physics detector
analysis \citep{johnson_celeritas_2021}.} which includes \ac{csg}, rectilinear
array, and hexagonal array trackers, leveraged via virtual functions. In Shift, GPU polymorphism has previously been avoided by 
supporting only a single universe/tracker type.

Prior to this work, Shift only supported the \ac{rtk} universe type on the GPU.
\Ac{rtk} is a special-purpose universe type that models a full reactor core,
and is separate from the \ac{orange} package. A single \ac{rtk} universe always
consists of (1) a rectilinear array representing the core, (2) rectilinear
arrays representing assemblies, and (3) cuboids containing concentric
cylinders, representing pins.  The \ac{rtk} tracker is optimized to track
neutrons through this specific configuration. As a result, the \ac{rtk} tracker
is expected to provide the best possible performance. However, \ac{rtk} has
limited applicability because not all fission reactors can be modeled with
nested rectilinear grids. Numerous reactor designs consist of hexagonal
assemblies \citep{habush_fsv_1968, hejzlar_terrapower_2013, betzler_tcr_2020},
and pebble bed reactors consist of irregular configurations of spherical fuel
elements \citep{andreades_kairos_2016, mulder_x-energy_2020}.  In addition,
\ac{mc} neutron and photon transport is used for other applications such as
nuclear fusion reactors \citep{juarez_iter_2021, kos_fusion_2023} and
accelerator devices \citep{radel_shine_2016, nelson_niowave_2022}, which
involve geometries with complexity far beyond rectilinear grids.

In this work, three methods for GPU-based multi-universe tracking were
implemented in \ac{orange} to assess the trade-offs between single and multiple
tracker types. The timing results for Shift $\keff$ calculations on the
Summit~\citep{summit} and Frontier~\citep{frontier} supercomputers at \ac{ornl}
were recorded for each of the three methods and compared to \ac{rtk} timing
results.  The three methods are outlined as follows. The dynamic polymorphism
(DP) method employs multiple tracker types with virtual functions.  The static
polymorphism (SP) method employs multiple tracker types with switch statements.
The single tracker (ST) method avoids polymorphism by using a single tracker
type for all universe types. The ST method is accomplished using
\emph{pseudo-array} universes (a term specific to this work), which are arrays modeled as \ac{csg} universes
and tracked upon with a \ac{csg} tracker. Although this approach forfeits the
aforementioned benefits of array trackers, this is counteracted with a \ac{bih}
acceleration structure \citep{wachter_bih_2006}.

The remainder of this work proceeds as follows. Section~\ref{sec:background}
provides background on how Shift solves for $\keff$, including the role of
tracking operations. Section~\ref{sec:tracking} introduces tracking algorithms
for \ac{csg} and array universes, multi-universe geometries, and \ac{rtk}
universes. Section~\ref{sec:methodology} describes the implementation of each
of the three experimental multi-universe GPU tracking methods.
Section~\ref{sec:hardware} describes the computer hardware on Summit and
Frontier used for this work. Section~\ref{sec:smr} provides a performance
comparison of the three methods using a model of the NuScale \ac{smr}
\citep{nuscale}, composed of rectilinear assemblies. Section~\ref{sec:empire}
demonstrates the flexibility of the ST method on the Empire microreactor
benchmark problem \citep{lee_empire_2020, matthews_empire_2021} composed of
hexagonal assemblies.

\section{Background}\label{sec:background}

This section provides background on how Shift solves for $\keff$.
Section~\ref{sec:neutron_transport_equation} describes the neutron transport
equation formulated with $\keff$ as the dominant eigenvalue.
Section~\ref{sec:power_iteration} describes the power iteration solution method
and the role of tracking operations in the \ac{mc} random walk process.
Section~\ref{sec:shift} describes the Shift implementation of \ac{mc} power
iteration on the GPU.

\subsection{Neutron transport equation}\label{sec:neutron_transport_equation}

Neutrons interact with fuel and non-fuel nuclei via a variety of different
mechanisms which are broadly organized into three categories:

\begin{enumerate}
\item absorption: a neutron is captured by a nucleus,
\item fission: a neutron is captured by a nucleus, which subsequently
      breaks apart,
\item scattering: a neutron interacts with a nucleus, effectively changing the neutron's
      kinetic energy and direction.
\end{enumerate}

As a neutron travels through a material, the probability that it will undergo
one of these interactions is quantified using nuclear \emph{cross sections},
functions which depend on the energy of the neutron, the types of nuclei within
the material, and the density and temperature of the material.  Absorption and
fission cross sections tend to be negatively correlated with neutron energy. As
a result, within a reactor core, neutrons typically scatter $\sim 10^1$ times
before being terminated via absorption/fission. These effects are quantified in
the neutron transport equation:

\begin{equation}\label{eq:transport}
(\op{T} -\op{S})\psi = \frac{1}{\keff}{\op{\chi}}\op{F}\psi\:,
\end{equation}

\noindent where $\op{T}$ is the transport operator, $\op{S}$ is the scattering
operator,  $\op{\chi}$ is the energy spectrum of neutrons born from fission,
$\op{F}$ is the fission operator, and $\psi$ is the neutron flux, which
describes the number of neutrons that pass through a 2D area per unit time per
unit angle. The $\op{T}$, $\op{S}$, and $\op{F}$ operators depend on nuclear
cross sections, but these relationships are omitted for brevity.  By collecting
the operators in a single term,

\begin{equation}
\op{A} \equiv \op{F}(\op{T}-\op{S})^{-1}{\op{\chi}}\:,
\end{equation}

\noindent Equation~\ref{eq:transport} can be formulated as a standard
eigenvalue problem,

\begin{equation} \label{eq:eigenvalue_problem}
\keff \fsrc = \op{A}\fsrc,
\end{equation}

\noindent where $\keff$ is the dominant eigenvalue of $\op{A}$ and $\fsrc$, the
eigenvector, is the fission source given by

\begin{equation}
\fsrc = \op{F} \psi.
\end{equation}

Physically, $\fsrc$ is a probability density function describing the spatial
distribution of fission neutrons. Like $\keff$, this distribution is not known
a priori. Due to the complexity of the geometry and cross sections,
Equation~\ref{eq:eigenvalue_problem} must be solved numerically for most
practical cases.

\subsection{Monte Carlo power iteration}\label{sec:power_iteration}

The eigenvalue problem presented in Equation~\ref{eq:eigenvalue_problem} can be
solved with \ac{mc} power iteration. With this method, $\keff$ and $\fsrc$ are
solved iteratively by applying the standard power iteration eigenvalue
algorithm \citep{mises_power_1929} to Equation~\ref{eq:eigenvalue_problem},

\begin{equation}\label{eq:pi_1}
\fsrc^{(n+1)} = \frac{1}{\keff^{(n)}} \op{A} \fsrc^{(n)},
\end{equation}
\begin{equation}\label{eq:pi_2}
\keff^{(n+1)} 
=  \keff^{(n)} \frac{\langle \fsrc^{(n+1)} \rangle}{\langle \fsrc^{(n)} \rangle},
\end{equation}

\noindent where $\langle \cdot \rangle$ denotes integration over space, energy,
and angle. This iteration scheme is carried out using
Algorithm~\ref{alg:power_iteration}, where \runhistories is an \ac{mc} neutron
transport simulation. As seen in Algorithm~\ref{alg:power_iteration},
iterations of Equations~\ref{eq:pi_1} and \ref{eq:pi_2}, referred to as
\emph{cycles}, are carried out in two stages.
First, a sufficient number of \emph{inactive} cycles are run to converge the
shape of $\fsrc$, with intermediate estimates of $\keff$ discarded.  Then
\emph{active} cycles are run, accumulating $\keff$ estimates to produce the
final $\keff$ with statistical uncertainty.
Statistical estimates of $\fsrc$ and other radiological quantities such as
neutron flux or specific reaction rates are also obtained from active cycles
(omitted from Algorithms~\ref{alg:power_iteration} and \ref{alg:mc} for
brevity). These estimates, often calculated within geometry cells or the volume elements of a superimposed
mesh, are called \emph{tallies}. Active cycles typically require longer compute
times than inactive cycles because of the additional overhead associated with
tallies.

\begin{algorithm}
\caption{\ac{mc} power iteration algorithm for calculating the converged
$\keff$ using $m_i$ inactive cycles, $m_a$
active cycles, and $n$ histories per cycle.}
\label{alg:power_iteration}
\begin{algorithmic}[1]
\Procedure{mc\_power\_iteration}{$m_i$, $m_a$, $n$}
\State Set $\fsrc$ to initial guess
\For{inactive\_cycle $\in [1, m_i]$}
    \State $\keff$, $\fsrc$ $=$ run\_histories($\fsrc$, $n$)
\EndFor
\For{active\_cycle $\in [1, m_a]$}
    \State $\keff$, $\fsrc$ $=$ run\_histories($\fsrc$, $n$)
    \State accumulate $\keff$
\EndFor
\State \Return average of all stored $\keff$
\EndProcedure
\end{algorithmic}
\end{algorithm}

The \runhistories \ac{mc} transport function used in
Algorithm~\ref{alg:power_iteration} is shown in Algorithm~\ref{alg:mc}. For
each of $n$ histories, a neutron birth position is sampled from the supplied
guess for the fission source distribution. A random walk technique is then used
to move the neutron through a computational representation of the reactor. This
algorithm makes use of the geometry tracking operations defined in
Table~\ref{table:tracking}.
Tracking operations typically account for 20\% or less of the total runtime
within an \ac{mc} simulation \citep{hamilton_gpu_2019}. Cross section
calculations make up the plurality of runtime because of the large number of
memory fetch operations. By definition, the random walk process imposes
divergent neutron paths across histories, resulting in random memory access. As
a result, \ac{mc} simulations tend to be latency bound.

\begin{algorithm}
\caption{\ac{mc} neutron transport algorithm for simulating $n$ neutron
histories born from $\fsrc$, in order to update estimates of $\keff$ and
$\fsrc$. This algorithm uses the geometric tracking operations specified in
Table~\ref{table:tracking}.}
\label{alg:mc}
\begin{algorithmic}[1]
\Procedure{run\_histories}{$\fsrc$, $n$}
\For{history $\in [1, n]$}
    \State sample position ($r$) from $\fsrc$
    \State sample energy ($E$) from fission energy spectrum
    \State sample direction ($\Omega$) isotropically
    \State cell ($c$) = \findcell\unskip($r$)
    \While{true}
        \State calculate the cross section ($\Sigma$) in $c$
        \State distance ($d$), surface ($s$) = \distancetosurface\unskip($r$, $\Omega$)
        \State sample $\#$ of mean free paths ($\tau$) before event
        \While{$d < \tau/\Sigma$}
            \State $\tau$ $=$ $\tau$ - $\Sigma$ $\times$ d
            \State $r$ = \movewithincell\unskip($r$, $\Omega$, $d$)
            \State $c$ = \crosssurface\unskip($r$, $c$, $s$)
            \State calculate $\Sigma$ in $c$
            \State $d$, $s$ = \distancetosurface\unskip($r$, $\Omega$)
        \EndWhile
        \State $r$ = \movewithincell\unskip($r$, $\Omega$, $\tau/\Sigma$)
        \State sample event type ($\xi$)
        \If{$\xi$ == scatter}
            \State update $E$ and $\Omega$
        \ElsIf{$\xi$ == fission}
            \State calculate and store an estimate of $\keff$
            \State store fission sites
            \State break 
        \ElsIf{$\xi$ == absorption}
            \State break 
        \EndIf
    \EndWhile
\EndFor
\State calculate updated $\keff$ from stored $\keff$ values
\State calculate updated $\fsrc$ from stored fission sites
\State \Return updated $\keff$, updated $\fsrc$
\EndProcedure
\end{algorithmic}
\end{algorithm}

\subsection{GPU-based implementation in Shift}\label{sec:shift}

On the CPU, Shift carries out \ac{mc} power iteration using the \ac{mc} neutron
transport simulation algorithm shown in Algorithm~\ref{alg:mc}. On the GPU,
Shift uses an \emph{event-based} \ac{mc} algorithm. As in
Algorithm~\ref{alg:mc}, the event-based algorithm involves simulating $n$
histories per cycle; however, operations are reordered to take advantage of
\ac{simt} parallelism. All operations---including birth, tracking operations,
and collisions---are performed on a vector of histories. This vector is masked
in order to only perform operations on applicable histories.  Each of the
tracking operations in Table \ref{table:tracking} is called for a vector of
histories via a kernel of the same name.  Full details of the event-based
transport algorithm are found in \cite{hamilton_gpu_2019}. This approach leads
to smaller kernel sizes and therefore increased occupancy and higher tracking
rates.

\section{Tracking algorithms}\label{sec:tracking}

This section discusses the implementation of the Table~\ref{table:tracking}
tracking operations for different universe types. For brevity, only the two
most algorithmically interesting tracking operations are discussed: \findcell
and \crosssurface.
Section~\ref{sec:single_universe_tracking} provides possible implementations of
these operations for \ac{csg} and rectilinear array universes, where the latter
are shown to have significantly lower time complexity.  Hexagonal array
universe tracking algorithms can be implemented using the same strategy as that
used for rectilinear arrays, with the same time complexity.  However, hexagonal
array indexing is considerably more complicated, so these algorithms are
omitted here for simplicity.
Section~\ref{sec:multi_universe_tracking} provides the implementation of
multi-universe tracking on the CPU within \ac{orange}.
Section~\ref{sec:rtk_tracking} provides the tracking algorithms for the
reactor-specific \ac{rtk} universe type, used by Shift on the GPU.

\subsection{Single universe tracking}
\label{sec:single_universe_tracking}

Possible implementations of \findcell and \crosssurface for a \ac{csg} tracker
are shown in Algorithms \ref{alg:find_cell_csg} and
\ref{alg:cross_surface_csg}.  Because \ac{csg} universes have no underlying
structure, \findcell involves conducting an $O(N)$ search over all of the cells
within the universe.  The \crosssurface implementation relies on a list of
\emph{neighbor} cells connected to each surface, which can be generated as a
preprocessing step. This neighbor list can then be searched in $O(N)$ time.

\begin{algorithm}
\caption{Possible \ac{csg} tracker version of \findcell.}
\label{alg:find_cell_csg}
\begin{algorithmic}[1]
\Procedure{find\_cell}{pos}
\For{cell $\in$ cells}
   \If{cell contains pos}
       \State \Return cell
   \EndIf
\EndFor
\EndProcedure
\end{algorithmic}
\end{algorithm}

\begin{algorithm}
\caption{Possible \ac{csg} tracker version of \crosssurface.} 
\label{alg:cross_surface_csg}
\begin{algorithmic}[1]
\Procedure{cross\_surface}{pos, cell, surf}
\For{new\_cell $\in$ neighbors(surf)}
   \If{new\_cell == cell}
       \State continue
   \EndIf
   \If{new\_cell contains pos}
       \State \Return new\_cell
   \EndIf
\EndFor
\EndProcedure
\end{algorithmic}
\end{algorithm}

Possible implementations of \findcell and \crosssurface for a rectilinear array
tracker are shown in Algorithms \ref{alg:find_cell_rect} and
\ref{alg:cross_surface_rect}. These algorithms have favorable time complexity
relative to their \ac{csg} counterparts. 
By conducting a binary search over the
mesh divisions, \findcell can be performed in $O(\log N)$ time.
\footnote{
For rectilinear grids with uniform spacing, \findcell can be performed in
$O(1)$ time by calculating the array indices directly in each dimension.
However, the presence of gaps between adjacent assemblies in most reactors
necessitates the use of non-uniform spacing and therefore a full binary search.
}
The \crosssurface operation is performed in $O(1)$ time
because the surface of each cell is known to have only one neighbor. It
requires the current cell, the $i/j/k$ dimension of the surface being crossed,
and whether the surface is being crossed in the positive or negative direction.

\begin{algorithm}
\caption{Possible rectilinear array tracker version of \findcell.}
\label{alg:find_cell_rect}
\begin{algorithmic}[1]
\Procedure{find\_cell}{pos}
\For{dim $\in [i, j, k]$}
    \State ijk[dim] = binary\_search(mesh[dim], pos[dim])  
\EndFor
    \State cell = ijk\_to\_cell(ijk)
    \State \Return cell
\EndProcedure
\end{algorithmic}
\end{algorithm}

\begin{algorithm}
\caption{Possible rectilinear array tracker version of \crosssurface.}
\label{alg:cross_surface_rect}
\begin{algorithmic}[1]
\Procedure{cross\_surface}{cell, dim, dir\_sign}
    \State ijk = cell\_to\_ijk(cell)
   \If{dir\_sign is positive}
       \State ijk[dim] = ijk[dim] + 1
   \Else
       \State ijk[dim] = ijk[dim] - 1
   \EndIf
    \State next\_cell = ijk\_to\_cell(ijk)
    \State \Return next\_cell
\EndProcedure
\end{algorithmic}
\end{algorithm}

\subsection{Multi-universe CPU tracking in ORANGE}
\label{sec:multi_universe_tracking}

The algorithms shown in Section~\ref{sec:single_universe_tracking} show how
possible implementations of \ac{csg} and array trackers operate within single universes. Here, tracking
algorithms for geometries consisting of multiple nested universes are shown, as
implemented on the CPU in \ac{orange}. The \ac{orange} CPU version of
\findcell and \crosssurface are shown in Algorithms~\ref{alg:find_cell_multi}
and \ref{alg:cross_surface_multi}. The \findcell algorithm recursively finds
the cell within daughter universes until reaching the bottom-most (i.e., most
embedded) level. In \crosssurface, a surf\_universe argument is supplied that
denotes the top-most (i.e., least embedded) level for which the neutron is on a
surface, noting that lower-level coincident boundary surfaces are removed during preprocessing.
The cell on the other side of the surface is then found by recursing through daughters
starting at this level.  Both algorithms assume the existence of a polymorphic
\gettracker function which returns a tracker object for a given universe,
depending on its type.

\begin{algorithm}
\caption{Multi-universe version of \findcell.}
\label{alg:find_cell_multi}
\begin{algorithmic}[1]
\Procedure{find\_cell}{pos}
\State tracker $=$ get\_tracker(root\_universe)
\State cell $=$ tracker.find\_cell(pos)
\While {cell.daughter}
    \State tracker $=$ get\_tracker(cell.daughter)
    \State cell $=$ tracker.find\_cell(pos)
\EndWhile
\State \Return cell
\EndProcedure
\end{algorithmic}
\end{algorithm}

\begin{algorithm}
\caption{Multi-universe version of \crosssurface. For simplicity, args can be
assumed to be a struct containing the union of the arguments to the standard
\ac{csg} and array versions of \crosssurface.}
\label{alg:cross_surface_multi}
\begin{algorithmic}[1]
\Procedure{cross\_surface}{surf\_universe, args}
\State tracker $=$ get\_tracker(surf\_universe)
\State next\_cell $=$ tracker.cross\_surface(args)
\While {next\_cell.daughter}
    \State tracker $=$ get\_tracker(next\_cell.daughter)
    \State next\_cell $=$ tracker.find\_cell(args.pos)
\EndWhile
\State \Return next\_cell
\EndProcedure
\end{algorithmic}
\end{algorithm}

\subsection{RTK tracking}\label{sec:rtk_tracking}

As mentioned in Section \ref{sec:introduction}, \ac{rtk} is the
reactor-specific universe type currently employed by Shift for GPU execution.
\Ac{rtk} is a template universe type explicitly instantiated to contain the three
nested levels of universes necessary to model reactors such as \acp{pwr} with
rectilinear configurations. An \ac{rtk} universe consists of a rectilinear
array core universe populated with rectilinear array assembly universes, each
populated with pin universes.  The pin universe is not a full, general-purpose
\ac{csg} universe, but rather a limited \ac{csg} universe consisting of
concentric cylinders within a rectangular cuboid cell.

The \ac{rtk} version of \findcell and \crosssurface are shown in
Algorithms~\ref{alg:find_cell_rtk} and \ref{alg:cross_surface_rtk}. These
algorithms resemble the standard multi-universe tracking algorithms shown in
Section~\ref{sec:multi_universe_tracking}, with two key differences.  First,
because there are exactly three levels of universes, the \emph{while} loops
over daughter universes can be unrolled (in practice, this is achieved through
\cpp template recursion). Second, the types of universes at each level are
known at compile time. Whereas the multi-universe tracking algorithms shown in
Section~\ref{sec:multi_universe_tracking} rely on a polymorphic \gettracker
function, Algorithms~\ref{alg:find_cell_rtk} and \ref{alg:cross_surface_rtk}
can call non-polymorphic \getrecttracker and \getcsgtracker functions.  As a
result of these simplifications, \ac{rtk} is expected to be the most performant
tracker type. However, the clear limitations on geometric complexity imposed by
\ac{rtk} motivate the implementation of general-purpose GPU tracking capabilities.

\begin{algorithm}
\caption{\Ac{rtk} version of \findcell.}
\label{alg:find_cell_rtk}
\begin{algorithmic}[1]
\Procedure{find\_cell}{pos}
\State core\_tracker $=$ get\_rect\_tracker(core\_universe)
\State core\_cell $=$ core\_tracker.find\_cell(pos)
\State assm\_universe = core\_cell.daughter
\State assm\_tracker $=$ get\_rect\_tracker(assm\_universe)
\State assm\_cell $=$ assm\_tracker.find\_cell(pos)
\State pin\_universe = assm\_cell.daughter
\State pin\_tracker $=$ get\_csg\_tracker(pin\_universe)
\State cell $=$ pin\_tracker.find\_cell(pos)
\State \Return cell
\EndProcedure
\end{algorithmic}
\end{algorithm}

\begin{algorithm}
\caption{\Ac{rtk} version of \crosssurface.}
\label{alg:cross_surface_rtk}
\begin{algorithmic}[1]
\Procedure{cross\_surface}{surf\_universe, args}
   \If{surf\_universe.level == core}
        \State core\_tracker $=$ get\_rect\_tracker(surf\_universe)
        \State core\_cell $=$ core\_tracker.cross\_surface(args)
        \State assm\_universe $=$ core\_cell.daughter
        \State assm\_tracker 
        \State \hspace{2em} $=$ get\_rect\_tracker(assm\_universe)
        \State assm\_cell $=$ assm\_tracker.find\_cell(args.pos)
        \State pin\_universe = assm\_cell.daughter
        \State pin\_tracker $=$ get\_csg\_tracker(pin\_universe)
        \State new\_cell $=$ pin\_tracker.find\_cell(args.pos)
   \ElsIf{surface\_universe.level == assembly}
        \State assm\_tracker $=$ get\_arr\_tracker(surf\_universe)
        \State assm\_cell $=$ assm\_tracker.find\_cell(args.pos)
        \State pin\_universe = assm\_cell.daughter
        \State pin\_tracker $=$ get\_csg\_tracker(pin\_universe)
        \State new\_cell $=$ pin\_tracker.find\_cell(args.pos)
   \ElsIf{surface\_universe.level = pin}
        \State pin\_tracker $=$ get\_csg\_tracker(surf\_universe)
        \State new\_cell $=$ pin\_tracker.find\_cell(args.pos)
   \EndIf
    \State \Return new\_cell
\EndProcedure
\end{algorithmic}
\end{algorithm}

\section{Methodology: Multi-universe GPU tracking methods}
\label{sec:methodology}

This section describes the \ac{orange} implementation of the three experimental
GPU-based multi-universe tracking methods explored in this work. The DP and SP
methods, which rely on universe-specific tracker types, were implemented only
for \ac{csg} and rectilinear array universes for the purposes of this work.
The ST method, which relies on only a single tracker, supports \ac{csg},
rectilinear array, and hexagonal array universe types.

\subsection{Dynamic polymorphism (DP) method}
\label{sec:dp_method}

This first method uses the standard multi-universe tracking algorithms put
forth in Section \ref{sec:multi_universe_tracking}, i.e., the approach used on
the CPU in \ac{orange}. A tracker base class is created, defining the tracking
operations listed in Table~\ref{table:tracking} as pure virtual methods.
\Ac{csg} and rectilinear array tracker classes inherit from the base class and
implement the virtual functions. A polymorphic \gettracker function returns a
pointer to either a \ac{csg} or rectilinear array tracker object. For a fair
comparison, within \ac{csg} universes, the DP method uses \ac{bih}
acceleration, which is described in Section \ref{sec:st_method}.

\subsection{Static polymorphism (SP) method}
\label{sec:sp_method}

Like the DP method, the SP method uses separate tracker types for \ac{csg} and
rectilinear array universes, but this is achieved with static polymorphism.
Using this approach, each tracker operation involves a switch statement
predicated on the type of the current universe (represented by an enumeration,
\texttt{UType}). For example, a possible implementation of \findcell is shown
in Listing~\ref{listing:find_cell_equiv}. This function takes three arguments:
the universe identifier (\texttt{uid}) of the current universe, the position,
and a set of geometry parameters which supply a mapping between \texttt{uid}
and \texttt{UType}. A \ac{csg} or rectilinear tracker is created based on the
\texttt{UType} of the universe specified by the \texttt{uid}.  The \findcell
method is then called on the tracker, and the resulting cell is returned.

\begin{minipage}{0.95\linewidth}
\begin{lstlisting}[caption={Possible \cpp implementation of \findcell using
static polymorphism. This code block has equivalent logic to the code in
Listings~\ref{listing:visit_universe}, \ref{listing:visit_tracker}, and
\ref{listing:find_cell}.}, label=listing:find_cell_equiv]
__host__ __device__ 
Cell find_cell(UniverseId uid,
               Position pos,
               Params params)
{
  switch(params.types[uid])
  {
    case UType::CSG:
        auto tracker
            = CSGTracker(uid, params)
        return tracker.find_cell(pos);
    case UType::RectArray:
        auto tracker
            = RectArrayTracker(uid,
                               params)
        return tracker.find_cell(pos);
  }
}

\end{lstlisting}
\end{minipage}

In \ac{orange}, this behavior is achieved via a template metaprogramming
approach which allows a single set of switch statement logic to be used by all
tracking operations, as shown in Listings~\ref{listing:visit_universe},
\ref{listing:visit_tracker}, and \ref{listing:find_cell}. In
Listing~\ref{listing:visit_universe}, a \texttt{Traits} struct is templated on
\texttt{UType}, and template specializations are used to define the
corresponding tracker type for each universe type. A
\texttt{visit\_universe\_type} function returns the output of a given functor,
which takes a \texttt{Traits} object as an argument.
Listing~\ref{listing:visit_tracker} provides a \texttt{visit\_tracker} function
which returns the output of a supplied functor \texttt{f}, which takes a
tracker of arbitrary type as an argument. This is achieved by wrapping
\texttt{f} in a second functor---\texttt{f2}, which calls \texttt{f} for a
given \texttt{Traits} object---and then calling \texttt{visit\_universe\_type}
with this second functor.

Listing~\ref{listing:find_cell} shows how the \texttt{visit\_tracker} function
can be used to achieve polymorphism. A \findcell functor is first created.
This functor, along with a \texttt{uid} and \texttt{params}, is passed
to \texttt{visit\_tracker}, and the resulting cell is returned.
Upon compilation, code in Listings~\ref{listing:visit_universe},
\ref{listing:visit_tracker}, and \ref{listing:find_cell} produce bitcode
equivalent to that of Listing \ref{listing:find_cell_equiv}.  The other
Table~\ref{table:tracking} tracking operations are implemented in an identical
fashion to the \findcell example in Listing~\ref{listing:find_cell}. As was the
case with the DP method, \ac{csg} universes in the SP method also use \ac{bih}
acceleration, which is described in Section~\ref{sec:st_method}.
Alternatives to this template metaprogramming approach include the curiously
recurring template pattern (CRTP) or \texttt{std::variant} with
\texttt{std::visit}, which would likely perform similarly.

\begin{minipage}{0.95\linewidth}
\begin{lstlisting}[caption={Simplified \cpp function for visiting different
                            universe types.},
                   label=listing:visit_universe]
template<UType U>
struct Traits;
template<>
struct Traits<UType::CSG>
{
  using tracker_type = CSGTracker;
}
template<>
struct Traits<UType::RectArray>
{
  using tracker_type = RectArrayTracker;
}

template<class F>
inline constexpr
__host__ __device__ decltype(auto)
visit_universe_type(F&& f, UType ut)
{
  switch (ut)
  {
  case UType::CSG:
    return f(Traits<UType::CSG>{})
  case UType::RectArray:
    return f(Traits<UType::RectArray>{})
  }
}
\end{lstlisting}
\end{minipage}

\begin{minipage}{0.95\linewidth}
\begin{lstlisting}[caption={Simplified \cpp function for visiting different
                            tracker types.},
                            label=listing:visit_tracker]
template<class F>
 __host__ __device__ decltype(auto)
visit_tracker(F&& f,
              UniverseId uid,
              Params params)
{
  auto f2 = [&](auto traits){
    return f(traits::tracker_type(
               uid,
               params));
    }

  return visit_universe_type(
           f2,
           params.universe_types[uid]);
}
\end{lstlisting}
\end{minipage}

\begin{minipage}{0.95\linewidth}
\begin{lstlisting}[caption={\cpp implementation of \findcell using the
                            \visittracker function shown in
                            Listing~\ref{listing:visit_tracker}.},
                            label=listing:find_cell]
auto find_cell = [&pos](auto&& tracker){ 
    return tracker.find_cell(pos);
    }
Cell cell =  visit_tracker(find_cell,
                           uid,
                           params)
\end{lstlisting}
\end{minipage}

\subsection{Single tracker (ST) method}
\label{sec:st_method}

While the DP and SP methods benefit from tracking algorithms optimized for
specific universe types, it is not clear that these benefits offset the
potential performance pitfalls of polymorphism, i.e., virtual function calls in
the case of the DP method and divergent execution paths in the case of the DP and SP
methods.
For comparison, instead of employing multiple polymorphic tracker types, the
final approach uses a single \ac{csg}
tracker for \ac{csg}, rectilinear array, and hexagonal array universes.  This
is accomplished by converting rectilinear and hexagonal array universes into
\ac{csg} universes. To do so, array cells are explicitly modeled as \ac{csg}
cells using the method shown in Figure~\ref{fig:csg}.  This conversion is done
automatically in \ac{orange}, and the resulting \ac{csg} universes are referred
to as ``pseudo-array universes,'' a term specific to this work. Modeling arrays
in this fashion is not a new approach: it is the simplest way of modeling an
array within an \ac{mc} code without specialized array universe types.

\begin{figure}
\centerline{\includegraphics[width=0.6\columnwidth]{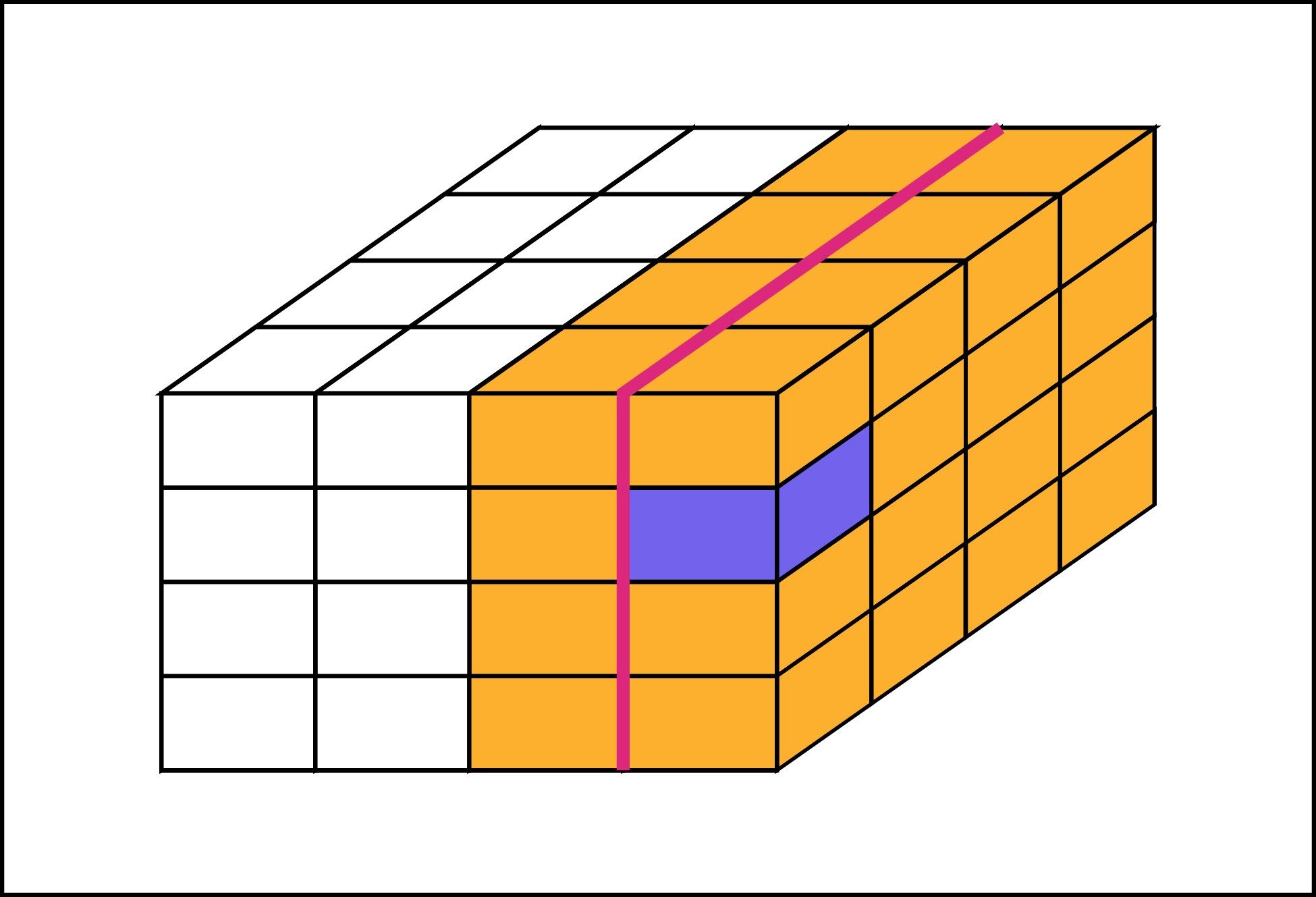}}
\caption{Example of a pseudo-rectilinear-array neighbor list. A neutron in the
purple cell crossing the magenta surface has all 31 gold cells as neighbors.}
\label{fig:neighbors}
\end{figure}

Because pseudo-array universes are \ac{csg} universes, they use \ac{csg}
tracking algorithms rather than the array versions that have improved time complexity. In other words,
for pseudo-arrays, naive implementations of \findcell
and \crosssurface would each use linear searches over all cells or all
neighbors, respectively.
The cost of the simple neighbor-based approach is exacerbated by the large
number of neighbor cells for pseudo-array surfaces,
since planar surfaces in \ac{csg} geometries are unbounded.
Figure~\ref{fig:neighbors} demonstrates the potential inefficiency: when crossing from
the purple cell through the magenta surface, which bounds all 32 colored cells,
the neighbor list includes all 31 gold cells.
It is noted that this issue does not arise with cell-based neighbor lists, an
alternative approach in which a mapping between each cell and its neighbor cells
is generated dynamically during runtime \citep{harper_neighbor_2020}.

\begin{figure*}
\centerline{\includegraphics[scale=0.8]{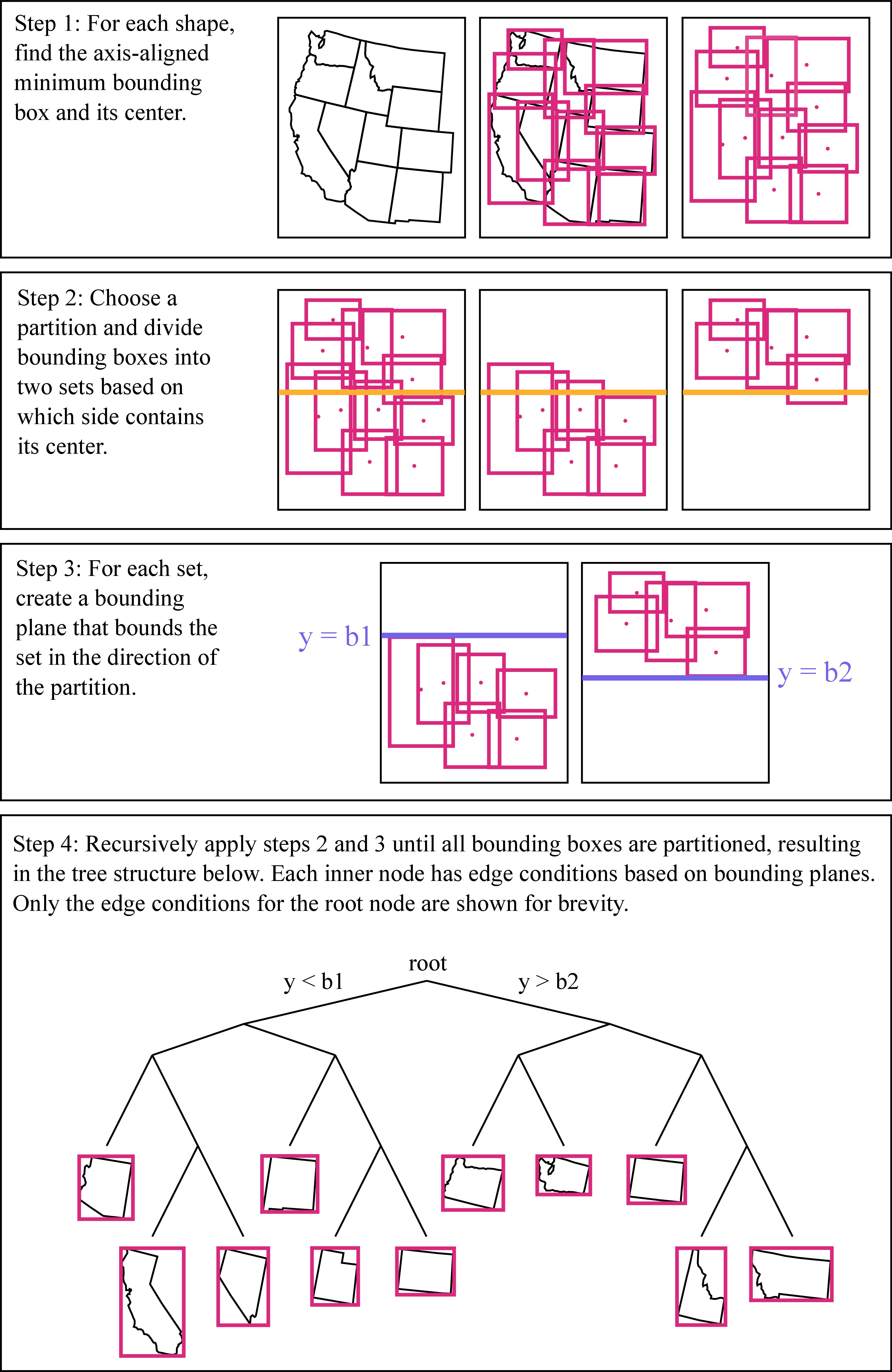}}
\caption{\Ac{bih} construction process demonstrated in 2D using the states of the
         American West. Map outline from \cite{map}.}
\label{fig:bih_construction}
\end{figure*}

To improve the performance of tracking operations for \ac{csg} universes,
including pseudo-array universes, a \ac{bih} acceleration structure was
implemented. The \ac{bih} construction process is demonstrated in
Figure~\ref{fig:bih_construction}.
First, axes-aligned bounding boxes are generated for each cell using a simple
method that involves successively truncating the universe bounding box using
the bounding planes associated with each of the cell's surfaces. Although the
nascent implementation of this method in \ac{orange} does not yet guarantee
minimum bounding boxes for arbitrarily complex cells, it is effective for the
simple geometric shapes found within the reactor models in this work. Once
bounding boxes are
ascertained, a partition plane is then chosen, and bounding boxes are
partitioned into two sets according to the location of each bounding box
center. For each of the two sets of partitioned bounding boxes, bounding planes
are created by moving the partition to fully enclose all bounding boxes. By
performing this process recursively on each set of bounding boxes, a binary
tree structure is created, with edges specifying the half-spaces which contain
all children.

One key feature of \ac{bih} trees is that for a given node, the half-spaces
created by the bounding planes may overlap. This guarantees that each cell
appears in the \ac{bih} tree exactly once, unlike $k$-D trees that must store a
cell twice if its bounding box is bisected by a partition.  This provides an
advantage in the pseudo-array use case in which adjacent cells may have
bounding boxes that overlap slightly because of floating point error.  \Ac{bih}
trees can handle this case without significant performance consequences.
\Ac{bvh} trees can also handle this overlap case. However, \ac{bvh} trees may
require more memory because they store six planes per node, significantly more
than the two planes required by \ac{bih} trees.  An open research question
within the \ac{bih} construction process is how to choose partitions. For the
purposes of this work, a standard \ac{sah} partitioning scheme
\citep{wald_sah_2007} was used.  This was implemented by evaluating three
equally spaced candidate partitions per axes at each partitioning step.
Candidate partitions were evaluated using a cost function balancing the number
of bounding boxes and the total surface area of bounding boxes appearing on
each side of the partition.

The \findcell and \crosssurface functions are accelerated by traversing the
\ac{bih} tree. For \findcell, traversal is terminated when a cell is found that
contains the supplied point. For \crosssurface, the traversal is terminated
when a cell is found that contains the supplied point, excluding the cell in
which the neutron originated. When traversing the \ac{bih} tree, both edge
conditions must be tested at each node because the half-spaces are allowed to
overlap.  \Ac{bih} trees are enabled for all \ac{csg} universes---not just
pseudo-array universes. As mentioned in Sections~\ref{sec:dp_method} and
\ref{sec:sp_method}, \ac{bih} trees are enabled for all \ac{csg} universes in
the DP and SP methods as well.

\section{Hardware description}\label{sec:hardware}

All simulations were performed on the Summit and Frontier supercomputers at
\ac{ornl}.  Summit has 4608 compute nodes, each consisting of two 22-core IBM
Power9 CPUs, and six NVIDIA Tesla V100 GPUs, each consisting of a single
\ac{gcd}. One core per Power9 is reserved for system tasks, leaving 42 usable
CPU cores per node. On the CPU side, 512 GB of RAM are available, and each V100
has 16 GB of RAM. Within this work, code was compiled on Summit with CUDA
11.5.2.

Frontier has 9408 compute nodes, each consisting of a 64-core AMD
$\text{3}^{\text{rd}}$ Gen EPYC CPU, and four AMD Radeon Instinct MI250X, each
consisting of two \acp{gcd}. Eight CPU cores are reserved for system tasks,
leaving 56 usable CPU cores per node. On the CPU side, Frontier also has 512 GB
of RAM, but each MI250X has 128 GB of RAM. As a result, Frontier has four times
as much GPU RAM per \ac{gcd} compared to Summit (64 GB vs. 16 GB). 
Within this work, code was compiled on Frontier with ROCm 5.6.0. 

\section{Performance testing}\label{sec:smr}

\begin{figure*}
\centerline{\includegraphics[scale=0.8]{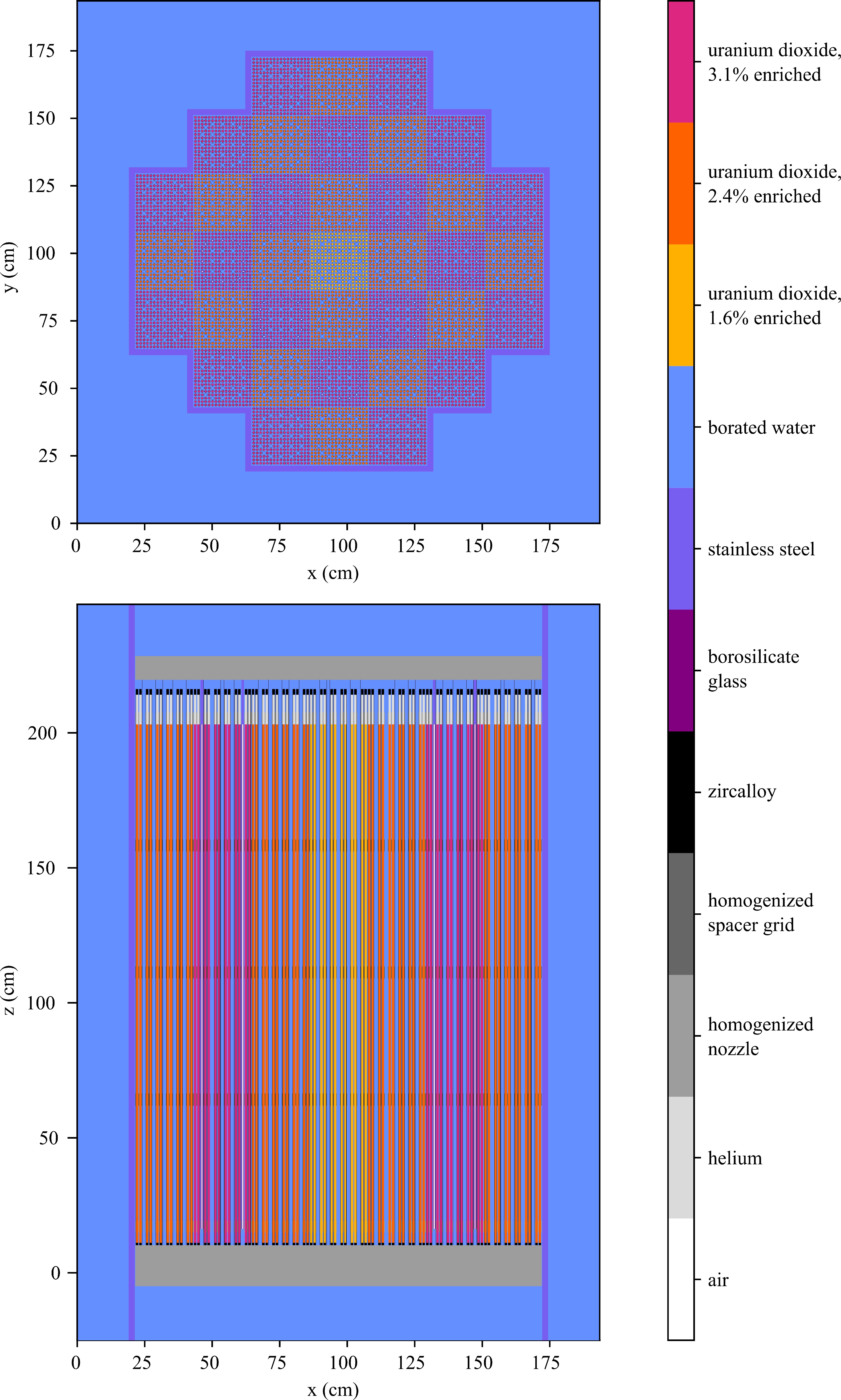}}
\caption{Midplane radial (top) and axial (bottom) slices of a full core model of
the NuScale \ac{smr} from \cite{smith_smr_2017}.}
\label{fig:smr_geom}
\end{figure*}

The performance of the three multi-universe GPU tracking methods described in
Section~\ref{sec:methodology} relative to \ac{rtk} (described in
Section~\ref{sec:rtk_tracking}) was assessed by obtaining timing results for a
rectilinear-array-based reactor problem. A full-core model of the NuScale
\ac{smr} \citep{smith_smr_2017}, shown in Figure~\ref{fig:smr_geom}, was chosen
for this purpose. The small size of the NuScale design---about one-eighth
the size of a typical \ac{pwr}---is economically attractive because of its low
capital cost \citep{black_smr_2019}, and also permits detailed full-core
\ac{mc} analysis. Likewise, this problem served as the challenge problem for
the ExaSMR project within the Exascale Computing Project, in which Shift was
used for GPU-based coupled neutron transport / thermal hydraulics analysis
\citep{merzari_gb_2023}.

The NuScale design consists of 37 assemblies arranged in a rectilinear grid.
Each assembly is a 17$\times$17 array of pins and contains uranium dioxide fuel
with a $^{235}$U enrichment of either 1.6\%, 2.4\%, or 3.1\% (by mass).  The
3.1\%-enriched fuel assemblies in the inner circle contain borosilicate glass
burnable neutron absorber rods. Spacer grids and nozzles have been homogenized
into slabs for simplicity. For this analysis, the fresh (i.e., non-depleted)
fuel version of the problem was used. Fresh fuel contains many fewer nuclides
than depleted fuel, thereby minimizing the time required to calculate cross
sections and maximizing the relative time spent on tracking operations.  As a
result, inactive cycles within this problem (where no time is spent on tallies)
represent the scenario in which tracking operations are expected to comprise
the largest fraction of runtime.

The computational representation of this reactor takes on several different
forms at runtime when testing the four tracking methods. With the \ac{rtk}
method, the entire core consists of a single \ac{rtk} universe. With the DP and
SP methods, the core consists of an array universe with embedded array
universes representing assemblies, each containing \ac{csg} universe pins.
\Ac{bih} acceleration is used within these \ac{csg} pin universes. This is
beneficial because each pin consists of a large number of cells ($\sim10^2$); due
to the complexity of the model, pins are split into 40 axial regions. With the
ST method, the core, assemblies, and pins are all \ac{csg} universes, and each
benefits from \ac{bih} acceleration.

For each of the four tracking methods, performance testing was conducted by
measuring the tracking rate, i.e., the number of histories simulated per unit
time, over a sweep of \emph{workloads}. Here, workload refers to the number of
histories per cycle assigned to each GPU \ac{gcd}.  Each trial consisted of 10
inactive cycles and 10 active cycles using all available GPU \acp{gcd} on a
single node of Summit or Frontier. During active cycles, the neutron flux
was tallied on a $119 \times 119 \times 30$ superimposed rectilinear mesh
(425,830 mesh volume elements). 

\begin{figure}
\centerline{\includegraphics[width=0.6\columnwidth]{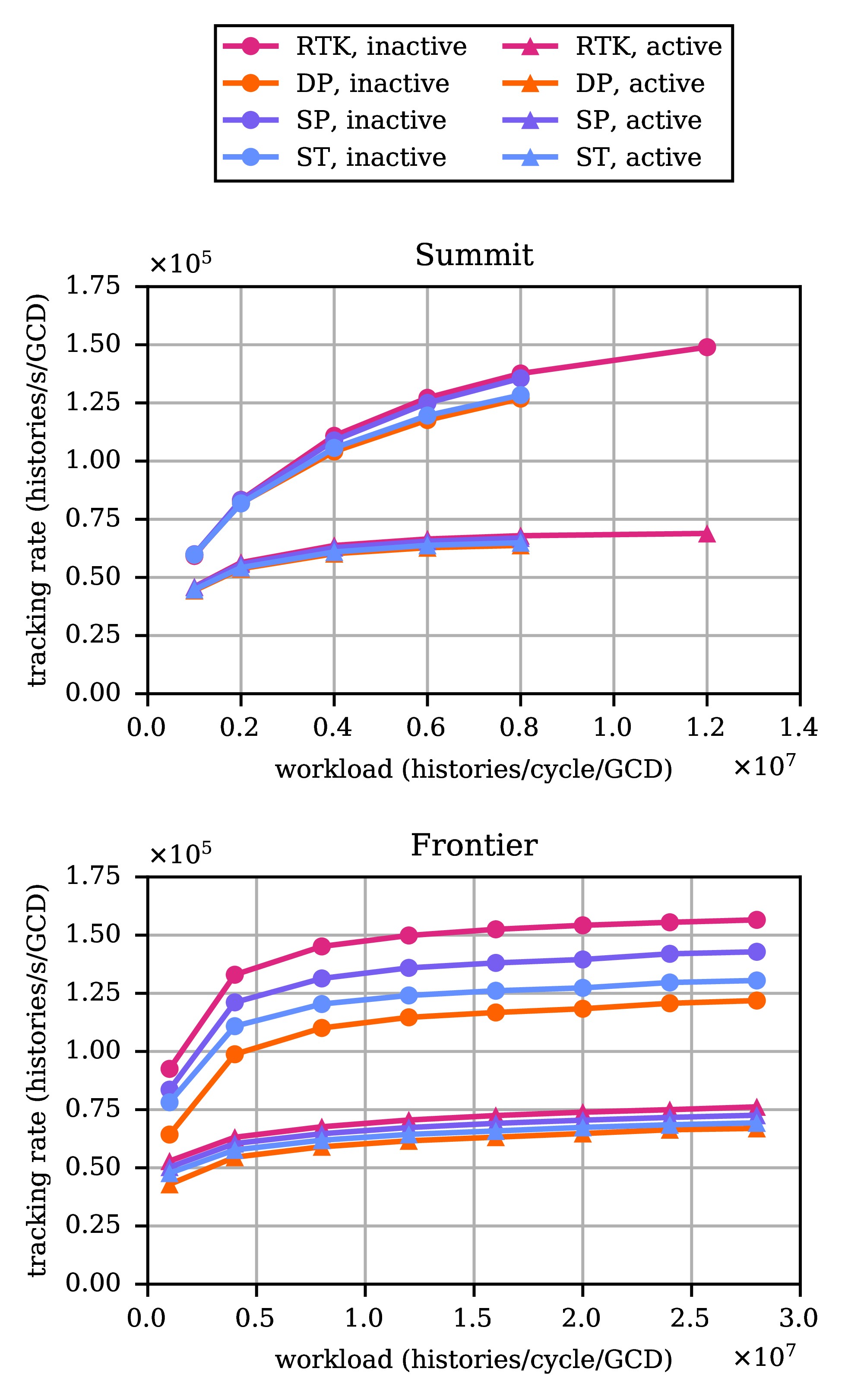}}
\caption{Neutron tracking rates as a function of the number of histories run by
each GPU \ac{gcd} on Summit and Frontier for the NuScale \ac{smr} problem.}
\label{fig:smr_both_wr}
\end{figure}

\subsection{Tracking rate results}\label{sec:smr_tracking}

Tracking rate results are shown in Figure~\ref{fig:smr_both_wr}, noting that
both plots have different $x$ scales but the same $y$ scale.  As expected, the
tracking rate increased with workload for all four methods, because large
workloads allow the GPU to more effectively hide the latency of memory fetches
associated with cross section calculations. Inactive cycle tracking rates are
higher because no time is spent processing tally results. On Frontier, tracking
rates for all four methods are near their asymptotic limit at a workload of
$2.8 \times 10^7$ histories/cycle/\ac{gcd}, well before running out of memory,
which was found to occur at $5\times 10^7$ histories/cycle/\ac{gcd} for
\ac{rtk} and  $3.7\times 10^7$ histories/cycle/\ac{gcd} for the DP, SP, and ST
methods.  However, on Summit, which has $4\times$ less RAM/\ac{gcd}, all four
methods ran out of memory prior to nearing an asymptotic limit, with \ac{rtk}
running out of memory at $1.2 \times 10^7$ histories/cycle/\ac{gcd} and the DP,
SP, and ST methods running out of memory at $8 \times 10^6$
histories/cycle/\ac{gcd}.

Aside from the $1 \times 10^6$ histories/s/\ac{gcd} inactive cycles, in which
the low workloads cause the kernel launch overhead to wash out any differences
between the methods, the relationships between the performance of the four
methods remained constant over all workloads. The highest tracking rates were
achieved with \ac{rtk}, as expected, followed by the SP method, the ST method,
and the DP method. Further analysis was performed with the $8 \times 10^6$
histories/cycle/\ac{gcd} trials on Summit and $2.8 \times 10^7$
histories/cycle/\ac{gcd} trials on Frontier. Tracking rates for each of the
four methods are shown in Table~\ref{table:tracking_rates}. Results show that
all three experimental methods achieve over 90\% of the \ac{rtk} tracking rate
on Summit. On Frontier, the DP, SP, and ST methods achieve 77.8\%, 91.2\%, and
83.4\% of the \ac{rtk} tracking rate, respectively, during inactive cycles, with all methods achieving
at least 87.9\% of the \ac{rtk} tracking rate during active cycles.
These results show that using universe-specific tracker types provides
better performance than using a single tracker type for all universe types,
provided that this can be done without virtual function calls.
However, although the SP method consistently provides the best performance, no method
incurred a significant performance penalty.

\begin{table*}
\small\sf\centering
\caption{GPU tracking rates for the NuScale \ac{smr} problem,
run on Summit with $8 \times 10^6$ histories/cycle/\ac{gcd}
and Frontier with $2.8 \times 10^7$ histories/cycle/\ac{gcd}.}
\label{table:tracking_rates}
\begin{tabular}{clcccc}
\toprule
 &          & \multicolumn{2}{c}{Summit}              & \multicolumn{2}{c}{Frontier} \\
 & Method   & GPU tracking       & Fraction of  & GPU tracking       & Fraction of   \\
 &          & rate               & RTK tracking & rate               & RTK tracking  \\
 &          & (histories/s/GCD)    & rate (\%)    & (histories/s/GCD)    & rate (\%)     \\
\midrule
\multirow{4}{*}{\rotatebox[origin=c]{90}{Inactive}}
 & RTK      & $1.38 \times 10^5$ &         100  & $1.57 \times 10^5$ & 100  \\
 & DP       & $1.27 \times 10^5$ &         92.2 & $1.22 \times 10^5$ & 77.8 \\
 & SP       & $1.36 \times 10^5$ &         98.5 & $1.43 \times 10^5$ & 91.2 \\
 & ST       & $1.28 \times 10^5$ &         93.3 & $1.31 \times 10^5$ & 83.4 \\
\midrule
\multirow{4}{*}{\rotatebox[origin=c]{90}{Active}}
 & RTK      & $6.79 \times 10^4$ &          100  & $7.62 \times 10^4$ & 100  \\
 & DP       & $6.38 \times 10^4$ &         93.9  & $6.69 \times 10^4$ & 87.9 \\
 & SP       & $6.69 \times 10^4$ &         98.6  & $7.26 \times 10^4$ & 95.3 \\
 & ST       & $6.49 \times 10^4$ &         95.5  & $6.93 \times 10^4$ & 91.0 \\
\bottomrule
\end{tabular}
\end{table*}

\subsection{Performance of individual tracking kernels}\label{sec:smr_kernels}

\begin{figure}
\centerline{\includegraphics[width=0.6\columnwidth]{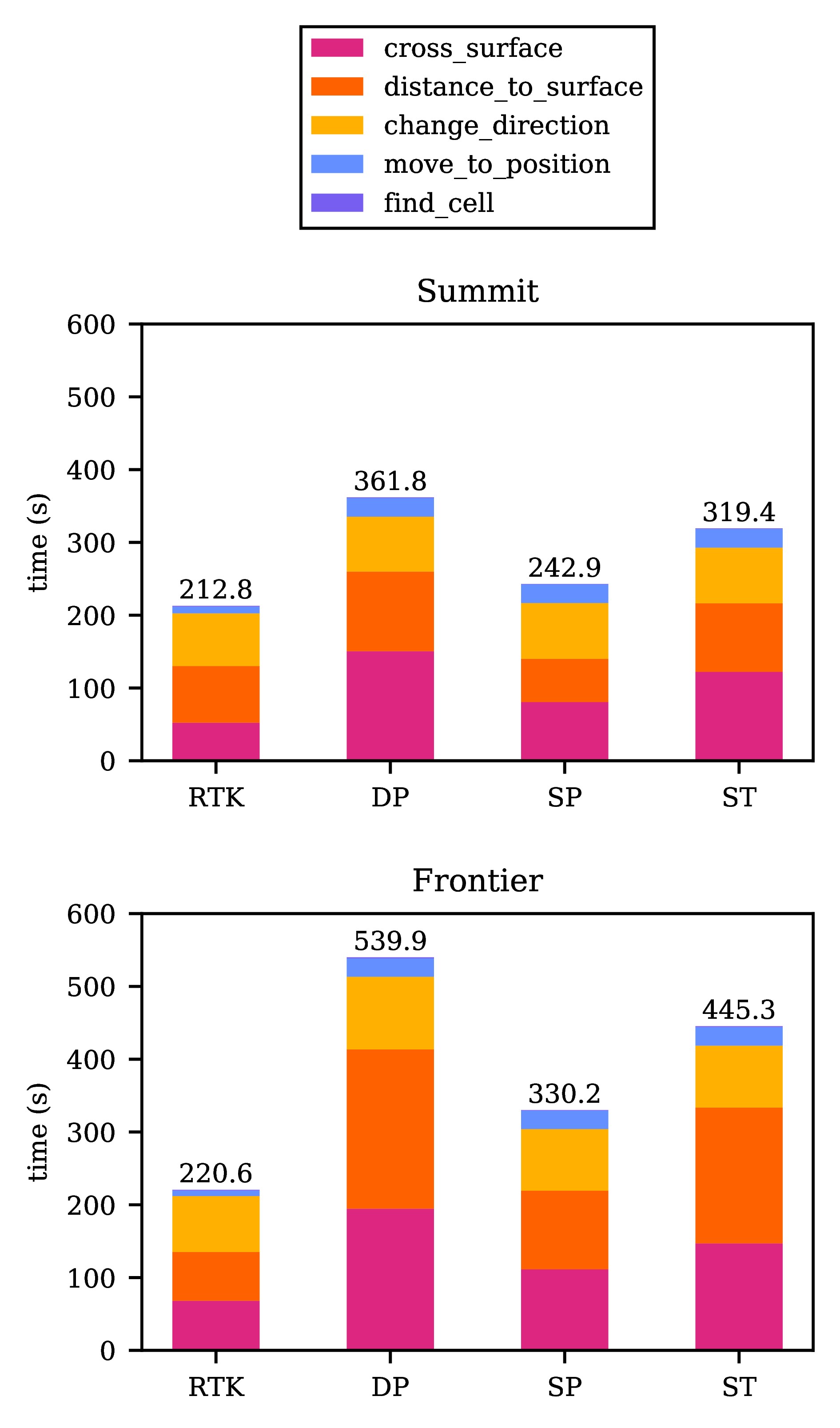}}
\caption{Total time spent in the 5 principal tracking kernels on Summit and
Frontier for the NuScale \ac{smr} problem, both with $8\times10^6$
histories/cycle/\ac{gcd}. All tracking kernels not listed here accounted for less
than 0.3\% of the total tracking time.}
\label{fig:smr_both_timers}
\end{figure}

The percentages of the total GPU runtime (inactive and active cycles) spent in
tracking kernels are shown in Table~\ref{table:percent}. These values vary from
12.1--26.1\%, indicating that tracking operations do not dominate runtime, as
expected.  Likewise, much larger differences in the performance of the four
methods are observed when only considering the time spent within tracking
kernels.  Figure~\ref{fig:smr_both_timers} shows the total time spent within
the five principal tracking kernels introduced in
Table~\ref{table:tracking}.  For a fair comparison, both Summit and
Frontier results are from workloads of $8\times10^6$ histories/cycle/\ac{gcd}.
This figure shows that the DP, SP, and ST methods spent 1.70$\times$,
1.14$\times$, and 1.50$\times$ more time, respectively, in tracking kernels compared to
\ac{rtk} on Summit, and 2.45$\times$, 1.5$\times$, and 2.02$\times$ more time, respectively,
compared to \ac{rtk} on Frontier. These results highlight the supremacy of the
SP method, almost achieving parity with \ac{rtk} on Summit.

\begin{table}
\small\sf\centering
\caption{Fraction of total GPU runtime (inactive and active cycles) spent in
geometry tracking kernels for the NuScale \ac{smr} problem.  Summit results are
for $8 \times 10^6$ histories/cycle/\ac{gcd} and Frontier results are for $2.8
\times 10^7$ histories/cycle/\ac{gcd}.}
\label{table:percent}
\begin{tabular}{lcc}
\toprule
         & \multicolumn{2}{c}{Runtime fraction (\%)} \\ 
         \cmidrule(lr){2-3} 
Method   & Summit & Frontier \\
\midrule
RTK      & 12.1   & 13.3 \\
DP & 19.2   & 26.1 \\
SP & 13.6   & 18.3 \\
ST & 17.2   & 23.3 \\
\bottomrule
\end{tabular}
\end{table}

\subsection{GPU to CPU performance comparison}\label{sec:smr_gpu_to_cpu}

Whereas overall tracking rates were slightly higher on Frontier, significantly
more time was spent in tracking kernels on Frontier compared to Summit.
Frontier also lags behind Summit when comparing GPU performance to CPU
performance. Table \ref{table:cpu_compare} compares GPU to CPU tracking rates
in terms of (1) the number of CPU cores required to match a single GPU
\ac{gcd}, and (2) the \emph{speedup}, defined as the ratio of the GPU tracking 
rate using all GPUs on a single node to the CPU tracking rate using all CPU cores on a single node.
Noting that both
GPU and CPU resources differ between these two machines, Summit has higher CPU
core equivalence and speedups compared to Frontier in all cases.  This can be
attributed to the fact that significantly better CPU performance is achieved on
Frontier, where a tracking rate of $4.94 \times 10^3$ histories/s/core was
achieved for inactive cycles, compared to $2.82 \times 10^3$ histories/s/core
for inactive cycles on Summit.

\begin{table}
\small\sf\centering
\caption{GPU to CPU performance comparison for the NuScale \ac{smr} problem,
run on Summit with $8 \times 10^6$ histories/cycle/\ac{gcd} and Frontier with
$2.8 \times 10^7$ histories/cycle/\ac{gcd}.
Speedup is defined as the ratio of the GPU tracking rate using all GPUs on a
single node to the CPU tracking rate using all CPU cores on a single node.
}
\label{table:cpu_compare}
\begin{tabular}{clcccc}
\toprule
 &          & \multicolumn{2}{c}{Summit}               & \multicolumn{2}{c}{Frontier} \\
                  \cmidrule(lr){3-4}                           \cmidrule(lr){5-6}
 & Method   & CPU core & Speedup    & CPU core    & Speedup \\
 &          & equivs.  &            & equivs.     &         \\
 &          & per GCD  &            & per GCD     &         \\
\midrule
\multirow{4}{*}{\rotatebox[origin=c]{90}{Inactive}}
 & RTK      & 48.9     & 6.98      & 31.7        & 4.53 \\
 & DP       & 45.0     & 6.43      & 24.7        & 3.53 \\
 & SP       & 48.1     & 6.87      & 28.9        & 4.13 \\
 & ST       & 45.6     & 6.51      & 26.4        & 3.78 \\
\midrule
\multirow{4}{*}{\rotatebox[origin=c]{90}{Active}}
 & RTK      & 28.1     & 4.01     & 15.3        & 2.19 \\
 & DP       & 26.3     & 3.76     & 13.4        & 1.92 \\
 & SP       & 27.7     & 3.95     & 14.6        & 2.08 \\
 & ST       & 26.8     & 3.83     & 13.9        & 1.99 \\
\bottomrule
\end{tabular}
\end{table}

\subsection{Profiling}\label{sec:smr_profiling}

In order to better understand the performance differences between the
four methods, profiling was done
using NVIDIA NSight Compute on a V100 (i.e., a Summit GPU) and AMD rocProf on an
MI250X (i.e., a Frontier GPU) using a workload of $8\times10^6$
histories/cycle/\ac{gcd}. Results appear in Table~\ref{table:profiling}. On both the
V100 and MI250X, RTK had the highest tracking rates and also the highest
theoretical occupancies for the two most expensive kernels (\crosssurface and
\distancetosurface).
The DP method had the lowest tracking rates on the V100 and MI250X and also the
lowest theoretical occupancies for \crosssurface and \distancetosurface on the V100.
On the V100, all kernels except \movewithincell had higher theoretical
occupancy with SP compared to DP, which was confirmed to be reflected
proportionally in register usage. The fact that the DP and SP methods are
identical other than their polymorphism implementation suggests that the virtual
functions themselves increase register usage and therefore decrease occupancy.
Achieved occupancy was observed to be strongly correlated with theoretical
occupancy and did not provide any further insights. On the MI250X, in contrast
to the V100, the DP method had the same theoretical occupancy as SP for all
kernels except \changedirection.

On both the V100 and MI250X, there were no differences in theoretical occupancy between SP
and ST because these two methods were run using the exact same kernels within the same executable. This
was done because the SP method still needs the ST code for \ac{bih}
acceleration within \ac{csg} universes. While the ST method does not need to be
compiled with the SP code in place, stripping out the relatively small and simple SP code did not result in
an increase in the ST tracking rate. Consequently, a single executable was maintained for simplicity.

On the V100, only minor differences were observed in the L1 and L2 cache hit
rate, and these were not correlated with tracking rate results. This is
expected because in all cases, cache performance is likely governed by cross
section lookups occurring within the physics kernel launches interspersed with
the geometry kernel launches. The differences in branch efficiency between
the methods were also not correlated with tracking rate results. Similar
results were obtained on the MI250X, noting that (1) only the L2 hit rate can be
obtained with rocProf, and (2) instead of branch efficiency, rocProf
provides vector arithmetic logic unit (VALU) utilization, which is correlated
with branch efficiency. 

Finally, on the V100, the number of warp stalls during instruction fetches correlated
strongly with tracking rate results.  \footnote{Obtained via the
\texttt{\seqsplit{smsp\_\_pcsamp\_warps\_issue\_stalled\_no\_instructions}}
metric.}
The DP method resulted in about $3\times$ as many warp stalls within the
\crosssurface and \distancetosurface kernels, suggesting that virtual function
calls put additional pressure on the instruction cache.

\begin{table*}
\centering
\caption{Profiling results on NVIDIA V100 and AMD MI250X GPUs using a workload of $8\times10^6$ histories/cycle/\ac{gcd}. Highlighted columns show the metrics correlated with performance results.
}
\label{table:profiling}
\begin{adjustbox}{angle=90}
\begin{tabular}{ll>{\columncolor{colcolor}}cccc>{\columncolor{colcolor}}c>{\columncolor{colcolor}}ccc}
\toprule
                      &                     & \multicolumn{5}{c}{V100}                                    & \multicolumn{3}{c}{MI250X} \\
                                               \cmidrule(lr){3-7}                                           \cmidrule(lr){8-10}  
Method                & Kernel              & Theoretical   & L1       & L2       & Branch       &  Warp stalls  & Theoretical& L2     & VALU        \\
                      &                     & occupancy     & hit      & hit      & efficiency   &  from         & occupancy  & hit    & utilization \\
                      &                     & (\%)          & rate     & rate     & (\%)         &  instruction  & (\%)       & rate   & (\%)        \\
                      &                     &               & (\%)     & (\%)     &              &  fetches      &            & (\%)   &             \\
                      &                     &               &          &          &              &   (\#)        &            &        &             \\
\midrule                                                                                                                            
\multirow{5}{*}{RTK}                                                                                                                
                      &  \crosssurface      & 75            & 57.8     & 71.9     & 89.9         & 657           & 100         &  70.4  &  86.5     \\
                      &  \distancetosurface & 50            & 58.0     & 66.5     & 85.9         & 326           & 100         &  53.7  &  89.6     \\
                      &  \changedirection   & 100           & 45.0     & 50.4     & 100          & 34            & 100         &  82.1  &  88.4     \\
                      &  \movewithincell    & 100           & 66.7     & 58.8     & 0.0          & 27            & 100         &  63.4  &  86.3     \\
                      &  \findcell          & 100           & 53.7     & 69.7     & 78.8         & 398           & 100         &  58.1  &  91.7     \\
\midrule                                                                                                                                
\multirow{5}{*}{DP}                                                                                                                     
                      &  \crosssurface      & 25            & 59.9     & 64.4     & 97.3         & 4851          & 50          &  69.4  &  87.6     \\
                      &  \distancetosurface & 12.5          & 59.6     & 67.8     & 89.6         & 10,031        & 50          &  54.5  &  89.6     \\
                      &  \changedirection   & 25            & 45.0     & 52.4     & 100          & 70            & 50          &  51.8  &  89.6     \\
                      &  \movewithincell    & 100           & 72.0     & 69.9     & 100          & 23            & 100         &  66.2  &  84.0     \\
                      &  \findcell          & 25            & 57.1     & 65.2     & 92.9         & 8796          & 62.5        &  58.5  &  91.7     \\
\midrule                                                                                                                                
\multirow{5}{*}{SP}                                                                                                                    
                      &  \crosssurface      & 37.5          & 68.3     & 72.7     & 97.9         & 1153          & 50          & 69.0   & 87.6      \\
                      &  \distancetosurface & 25            & 72.4     & 48.9     & 89.4         & 3741          & 50          & 54.5   & 89.6      \\
                      &  \changedirection   & 50            & 44.1     & 50.3     & 100          & 36            & 100         & 52.3   & 89.6      \\
                      &  \movewithincell    & 100           & 72.0     & 69.9     & 100          & 13            & 100         & 65.4   & 84.0      \\
                      &  \findcell          & 50            & 63.8     & 75.4     & 92.0         & 3710          & 62.5        & 58.4   & 91.7      \\
\midrule                                                                                                                                
\multirow{5}{*}{ST}                                                                                                                     
                      &  \crosssurface      & 37.5          & 70.3     & 74.1     & 98.7         & 6110          & 50         &  69.2  & 89.2      \\
                      &  \distancetosurface & 25            & 71.7     & 59.5     & 87.8         & 6523          & 50         &  54.4  & 89.6      \\
                      &  \changedirection   & 50            & 41.3     & 51.6     & 100          & 55            & 100        &  51.5  & 89.6      \\
                      &  \movewithincell    & 100           & 71.95    & 69.8     & 100          & 45            & 100        &  65.7  & 84.0      \\
                      &  \findcell          & 50            & 66.2     & 79.2     & 94.0         & 9995          & 62.5       &  58.2  & 91.7      \\
\bottomrule
\end{tabular}
\end{adjustbox}
\end{table*}

\subsection{Verification}\label{sec:smr_verfication}

\begin{figure*}
\centerline{\includegraphics[scale=0.8]{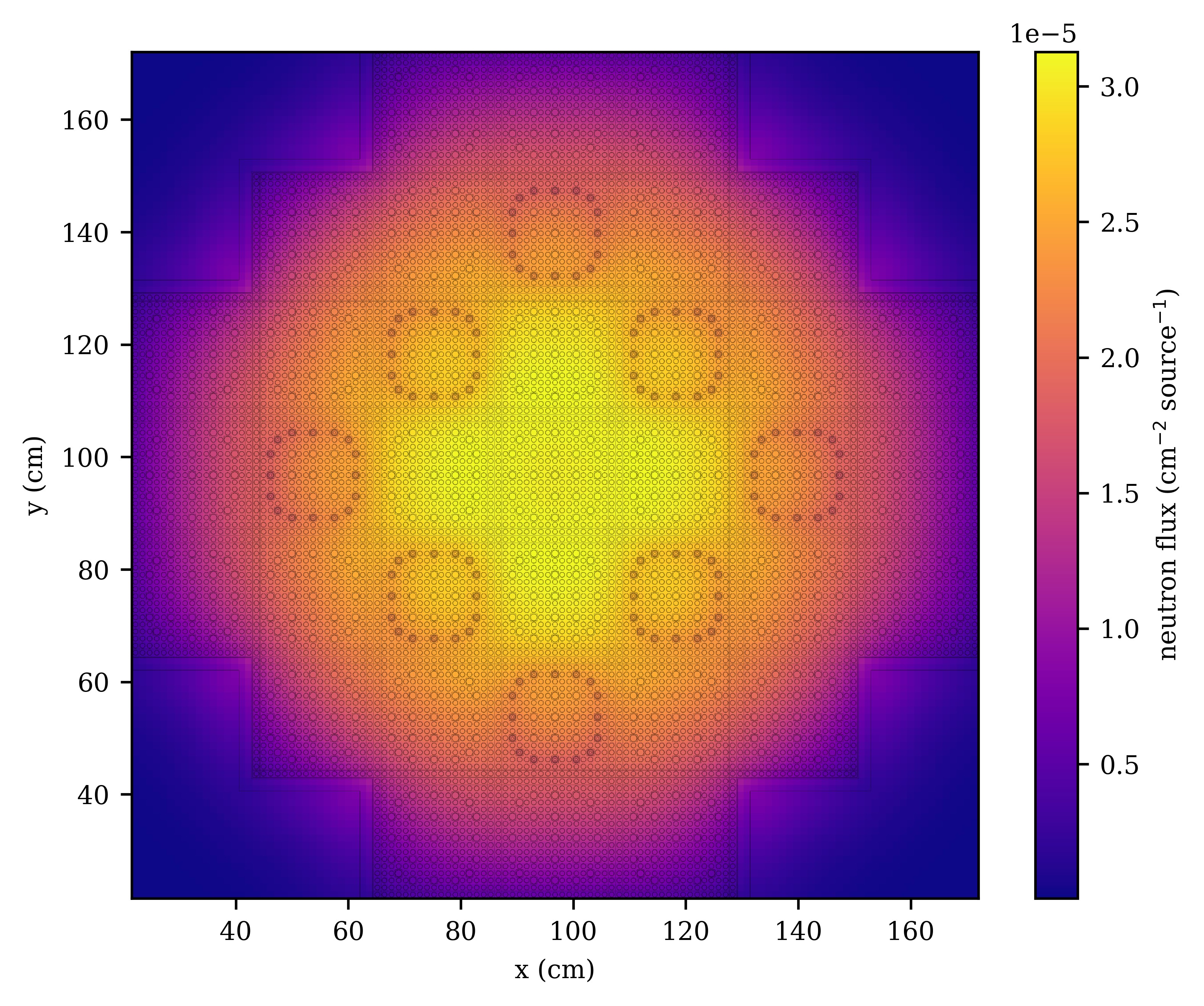}}
\caption{Converged neutron flux distribution on the midplane of the NuScale
\ac{smr} problem using the SP method, obtained on 50 nodes of Frontier with
$2.5 \times 10^5$ histories/cycle/\ac{gcd}, with 200 inactive and 400 active
cycles. The 1$\sigma$ statistical uncertainties in the flux on this slice are
all less than 0.15\% within the core region.}
\label{fig:smr_flux}
\end{figure*}

Single node performance testing trials, which were run with only 10 inactive
and 10 active cycles, were not sufficient to produce converged $\keff$ results.
To verify that all four tracking methods produce the same solution, a final
trial was performed for each method on 50 nodes of Frontier with a workload of
$2.5 \times 10^5$ histories/cycle/\ac{gcd}, for a total of $10^8$ histories per
cycle.  For each trial, 200 inactive cycles and 400 active cycles were run.  A
lower workload was necessary to complete all 600 cycles within Frontier's
walltime limit. Converged $\keff$ values agreed closely, as shown in
Table~\ref{table:keff}. Neutron flux results for all 4 methods also matched
expectations. Figure~\ref{fig:smr_flux} shows the converged flux for the SP
method trial. This figure clearly shows the depression on the flux within the
borosilicate glass rods, which act as neutron absorbers.

\begin{table}
\small\sf\centering
\caption{Converged $\keff$ values and 1$\sigma$ statistical uncertainties for
the NuScale \ac{smr} problem, obtained on 50 nodes of Frontier with  $2.5
\times 10^5$ histories/cycle/\ac{gcd}, with 200 inactive and 400 active
cycles.}
\label{table:keff}
\begin{tabular}{lc}
\toprule
Method   & $\keff$                \\
\midrule
RTK      & 1.046076 $\pm$ 0.000005 \\
DP       & 1.046076 $\pm$ 0.000005  \\
SP       & 1.046067 $\pm$ 0.000005 \\
ST       & 1.046072 $\pm$ 0.000005 \\
\bottomrule
\end{tabular}
\end{table}

\section{Demonstration problem}\label{sec:empire}

\begin{figure*}
\centerline{\includegraphics[scale=0.8]{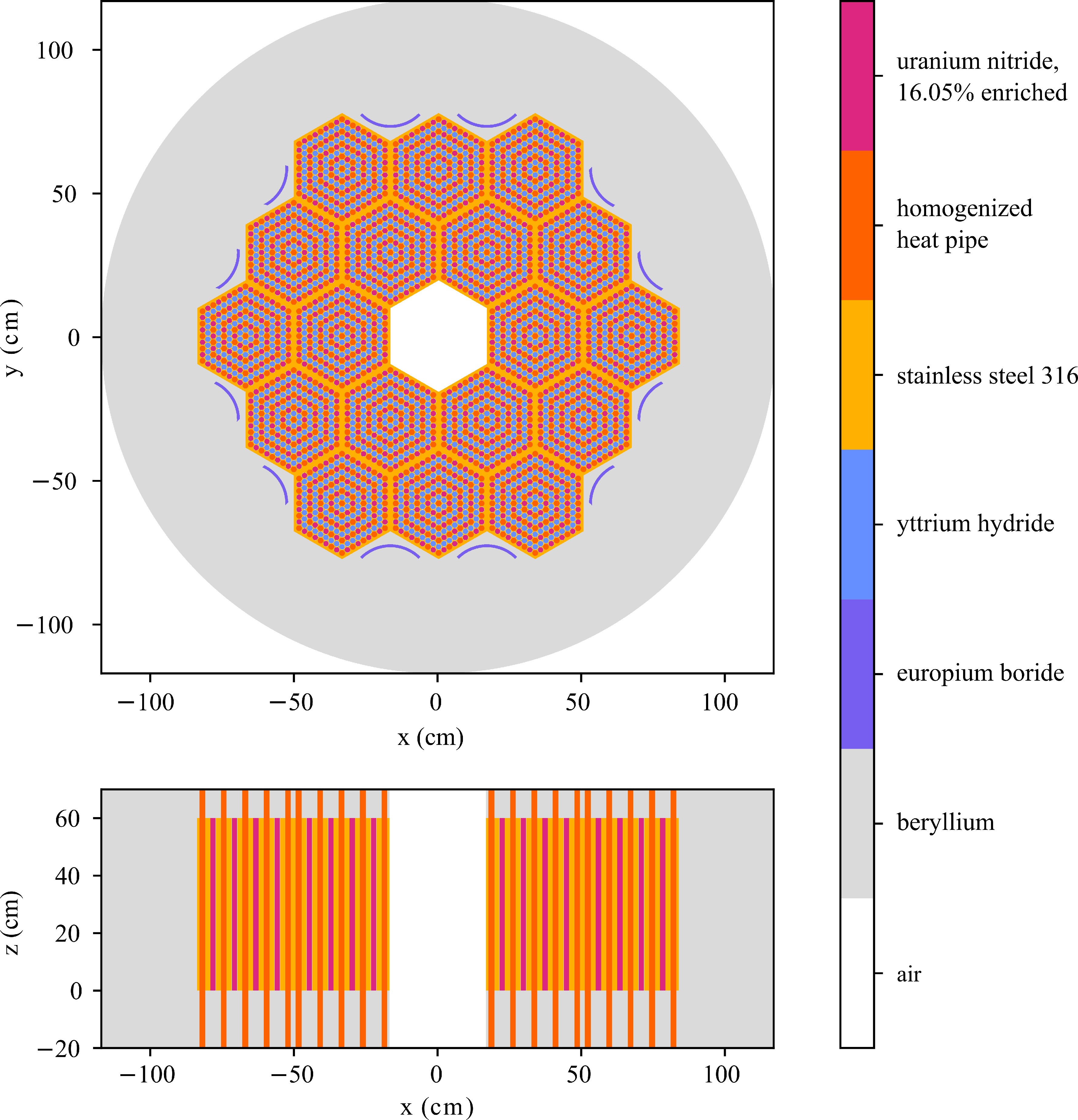}}
\caption{Midplane radial (top) and axial (bottom) slices of a full core model of
the Empire microreactor in ``drums-in'' configuration.}
\label{fig:empire_geom}
\end{figure*}

In Section \ref{sec:smr_tracking}, the SP method consistently provided the best
performance, but all three methods achieved tracking rates reasonably close to
that of \ac{rtk}. From a software engineering perspective, when considering
multiple approaches, the trade-off between performance and other factors such
as code simplicity and maintainability must be assessed.  The ST method
achieves over 90\% of the SP method tracking rate in all cases shown in Table
\ref{table:tracking_rates} and has the added benefit of not requiring an
additional hexagonal array tracker.  This is advantageous for
hexagonal-array-based reactors, which cannot be represented using \ac{rtk}, and
would otherwise require the development of a GPU-based hexagonal array tracker
for use with the SP method.  To demonstrate this benefit, the ST method was
used to perform a $\keff$ calculation on the Empire microreactor benchmark
problem that consists of nested hexagonal arrays, as shown in
Figure~\ref{fig:empire_geom}.  Microreactors, even smaller than \acp{smr}, are
designed to be factory assembled and easily transported in order to supply
power for remote areas, disaster relief, and space applications.

The Empire microreactor consists of 18 hexagonal assemblies arranged around a
central void region. Each assembly contains a total of 217 pins: 60 uranium
nitride fuel pins with 16.05\% $^{235}$U enrichment (by mass), 96 yttrium
hydride moderator pins, and 61 stainless steel heat pipes filled with liquid
sodium.  Within this model, material within the heat pipe pins is homogenized
for simplicity. The core is surrounded by 12 control drums which contain a
europium boride neutron absorber on one side. These drums can be rotated to
change the absorber configuration to control the reactivity within the core.
For this analysis, the ``drums-in'' configuration was used, with all absorbers
directly facing the core.

A Cartesian mesh tally would require a prohibitively large number of elements
in order to conform to the hexagonal pins and assemblies found in the
Empire geometry, and Shift does not yet support hexagonal mesh tallies on the
GPU. Thus, cell-based tallies were used to tally the flux and fission source
density within each fuel pin. For high-resolution results, each fuel pin was
subdivided into 360 cells, with 3 radial divisions, 4 circumferential
divisions, and 30 axial divisions. This resulted in a total of 388,800 cell
tallies, similar to the number of elements in the Cartesian mesh tally used for
the NuScale \ac{smr} problem.

The problem was run on 100 Frontier nodes using all eight \acp{gcd} on each
node, with a workload of $10^6$ histories/cycle/\ac{gcd}, for a total of
$8\times10^8$ histories/cycle. To achieve converged results, 120 inactive
cycles and 120 active cycles were performed. For comparison, CPU results were
obtained on a single node of Frontier with $2\times10^6$ histories/cycle for
120 inactive cycles and 120 active cycles.

\begin{figure*}
\centerline{\includegraphics[scale=0.9]{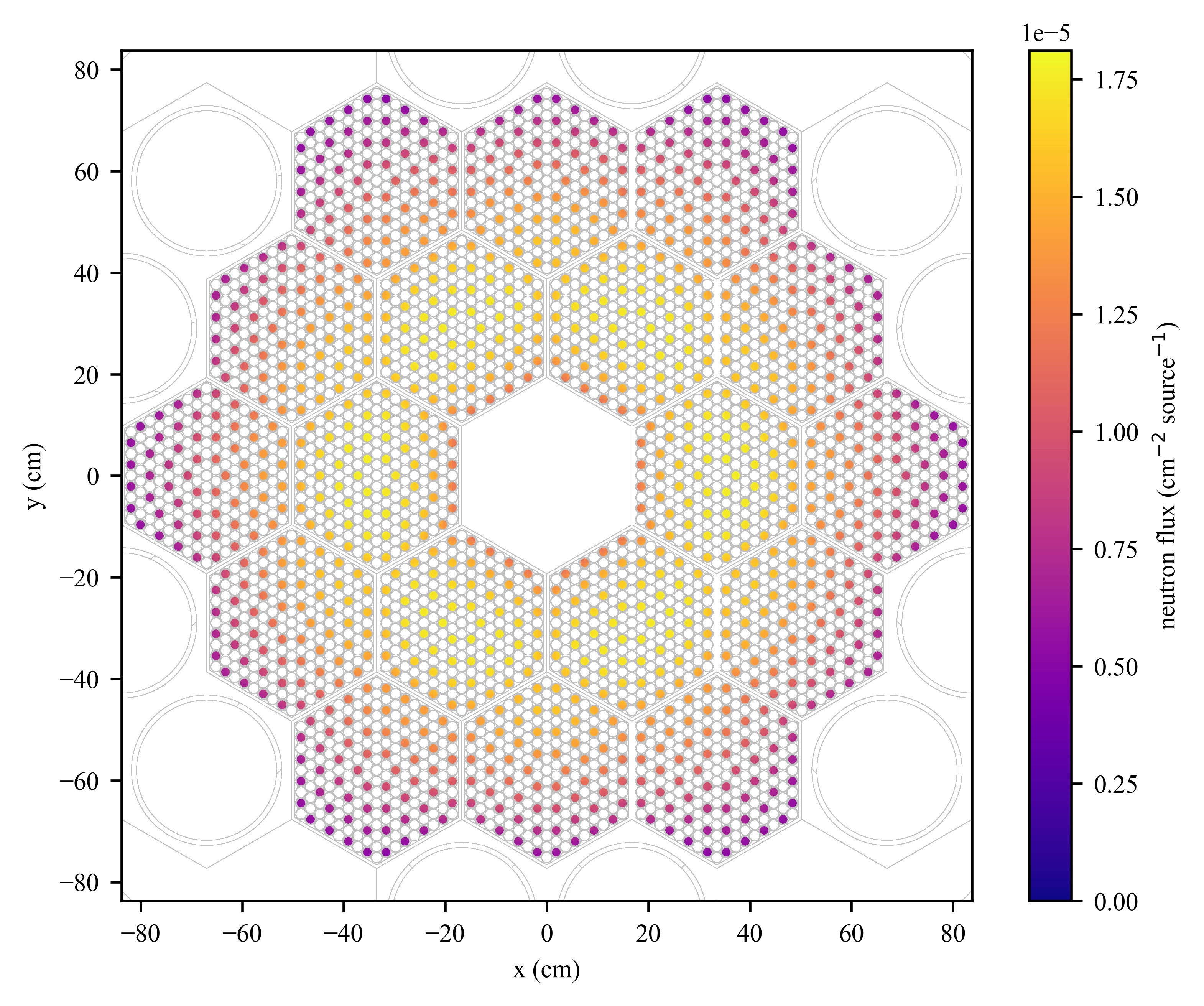}}
\caption{Converged neutron flux distribution on the midplane of the Empire problem
using the ST method, obtained on 100 nodes of Frontier with $10^6$
histories/cycle/\ac{gcd}, with 120 inactive and 120 active cycles. Statistical
uncertainties on this slice are all less than 0.10\%.}
\label{fig:empire_flux}
\end{figure*}

\begin{figure*}
\centerline{\includegraphics[scale=0.875]{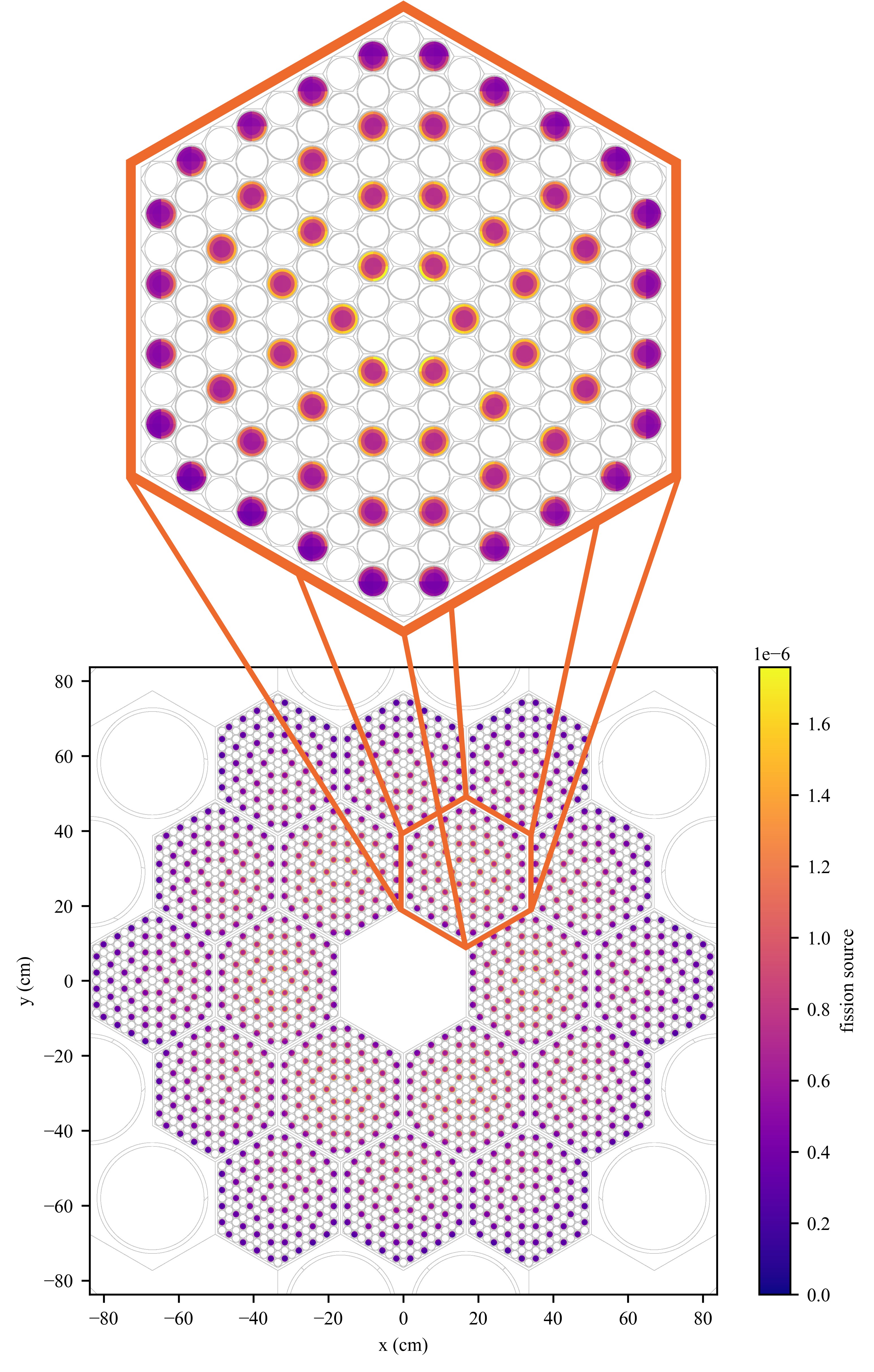}}
\caption{Converged fission source distribution on the midplane of the Empire
problem using the ST method, obtained on 100 nodes of Frontier with $10^6$
histories/cycle/\ac{gcd}, with 120 inactive and 120 active cycles. Statistical
uncertainties on this slice are all less than 0.25\%.}
\label{fig:empire_fission_combined}
\end{figure*}

Converged flux results are shown in Figure~\ref{fig:empire_flux}. As expected,
the highest flux occurs in an annular region around the center of the reactor, i.e.,
the region where leakage to the outside of the reactor and inner void
region are minimized. Figure~\ref{fig:empire_fission_combined} shows the
converged fission source.  As expected, the highest fission source density
occurs around the edges of fuel rods. For rods in the outer parts of
assemblies, the fission source density is observed to be highest on the inward-facing
edge. A converged $\keff$ of 1.026772 $\pm$ 0.000003 was obtained.

Table~\ref{table:empire_cpu_compare} shows tracking rates, as well as GPU vs.
CPU comparison results, in the same format as that used in
Table~\ref{table:cpu_compare}.
For inactive cycles, the tracking rate of $4.74 \times 10^4$ histories/s/\ac{gcd} is 36.2\% of the ST method tracking rate for
the NuScale \ac{smr} problem on Frontier.
For active cycles, the tracking rate of $4.33 \times 10^4$ histories/s/\ac{gcd} is 62.5\% of the ST method tracking rate for
the NuScale \ac{smr} problem on Frontier.
The 8.65\% degradation in tracking rate between inactive and active cycles for this problem is much less than the 47.1\% degradation observed in the NuScale \ac{smr} problem.
This is due to the fact that the cell tallies used in this problem have significantly less overhead than tracking on the superimposed mesh tally for the NuScale \ac{smr} problem.
When considering both inactive and active cycles, an overall speedup of 2.19$\times$ was obtained.

\begin{table}
\small\sf\centering
\caption{GPU to CPU performance comparison for the Empire problem on Frontier,
with $10^6$ histories/cycle/\ac{gcd}. 
Speedup is defined as the ratio of the GPU tracking rate using all GPUs on a
single node to the CPU tracking rate using all CPU cores on a single node.
}
\label{table:empire_cpu_compare}
\begin{tabular}{lccc}
\toprule
Cycle    & GPU tracking       & CPU core & Speedup  \\
type     & rate               & equivs.  &          \\
         & (histories/s/GCD)    & per GCD  &          \\
\midrule
Inactive & $4.74 \times 10^4$ & 15.9     & 2.27     \\
Active   & $4.33 \times 10^4$ & 14.8     & 2.12     \\
\bottomrule
\end{tabular}
\end{table}

\section{Conclusion}

In this work, three methods for GPU-based neutron tracking within
multi-universe geometries were tested and compared to \ac{rtk}.  For the
NuScale \ac{smr} problem, the DP, SP, and ST methods spent 2.45$\times$, 1.50$\times$,
and 2.02$\times$ more time, respectively, in tracking kernels in inactive cycles on Frontier
compared to \ac{rtk}. However, although inactive cycles on this fresh fuel
problem represent the maximum time spent performing tracking operations, these
methods still achieved 77.8\%, 91.2\%, and 83.4\%, respectively, of the \ac{rtk} tracking rate
on Frontier.  On the NVIDIA V100 (i.e., a Summit GPU), it appears that
performance differences between the methods can be explained by differences in device occupancy and
pressure on the instruction cache. On the AMD MI250X (i.e., a Frontier
GPU), occupancy was more weakly correlated to performance.
It is concluded that any of these three methods---which unlike RTK can handle
arbitrarily nested configurations---are suitable for production-level use,
provided that the fraction of runtime required by tracking operations is
consistent with the typical fission problems explored in this work (around
20\%).  For problems with geometric complexity far beyond these cases (perhaps
outside of fission applications), tracking performance differences become more
important, and the DP method should be avoided as it offers neither the
performance of the SP method nor the convenience of the ST method.
Broadly speaking, these results indicate that polymorphism can still be
effectively employed on the GPU, provided that it can be done without virtual
function calls. Improvements to GPU compilers may eliminate this limitation.

The SP method outperformed the DP and ST methods in nearly all cases.  However,
the ST method achieved over 90\% of the SP method's tracking rate in all cases,
and unlike the SP method, it requires only a single \ac{csg} tracker. The
flexibility of this approach was demonstrated with the Empire microreactor, a
hexagonal array-based reactor which cannot be represented using \ac{rtk}, and
would require an additional GPU-based hexagonal array tracker to be written in
order to use the SP method.  
On Frontier, the ST method was used to produce converged
$\keff$, neutron flux, and fission source results, and
an overall speedup of 2.19$\times$ over CPU execution was obtained.
This work will
facilitate GPU-based MC transport for reactor problems with geometric complexity
beyond rectilinear arrays, as well as other non-reactor radiation transport
problems with complex nested geometries.

\begin{funding}
This research was supported by the Exascale Computing Project (ECP), project
number 17-SC-20-SC. The ECP is a collaborative effort of two DOE organizations,
the Office of Science and the National Nuclear Security Administration, that
are responsible for the planning and preparation of a capable exascale
ecosystem---including software, applications, hardware, advanced system
engineering, and early testbed platforms---to support the nation's exascale
computing imperative.
\end{funding}

\begin{acks}
The authors would like to thank Stuart Slattery and Friederike Bostelmann for
their internal technical reviews.
\end{acks}


\begin{thebibliography}{32}
\providecommand{\natexlab}[1]{#1}
\providecommand{\url}[1]{\texttt{#1}}
\providecommand{\urlprefix}{URL }
\expandafter\ifx\csname urlstyle\endcsname\relax
  \providecommand{\doi}[1]{DOI:\discretionary{}{}{}#1}\else
  \providecommand{\doi}{DOI:\discretionary{}{}{}\begingroup
  \urlstyle{rm}\Url}\fi

\bibitem[{Andreades et~al.(2016)Andreades, Cisneros, Choi, Chong, Fratoni,
  Hong, Huddar, Huff, Kendrick, Krumwiede, Laufer, Munk, Scarlat and
  Zweibau}]{andreades_kairos_2016}
Andreades C, Cisneros AT, Choi JK, Chong AYK, Fratoni M, Hong S, Huddar LR,
  Huff KD, Kendrick J, Krumwiede DL, Laufer MR, Munk M, Scarlat RO and Zweibau
  N (2016) Design summary of the {M}ark-{I} pebble-bed, fluoride salt-cooled,
  high-temperature reactor commercial power plant.
\newblock \emph{Nuclear Technology} 195(3): 223--238.
\newblock \doi{10.13182/NT16-2}.

\bibitem[{Bentley(1975)}]{bentley_kd_1975}
Bentley JL (1975) Multidimensional binary search trees used for associative
  searching.
\newblock \emph{Communications of the Association for Computing Machinery}
  18(9): 509--517.
\newblock \doi{10.1145/361002.361007}.

\bibitem[{Betzler et~al.(2020)}]{betzler_tcr_2020}
Betzler BR et~al. (2020) {T}ransformational {C}hallenge {R}eactor preliminary
  core design report.
\newblock Technical Report {ORNL/TM-2020/1718}, Oak Ridge National Laboratory.

\bibitem[{Black et~al.(2019)Black, Aydogan and Koerner}]{black_smr_2019}
Black GA, Aydogan F and Koerner CL (2019) Economic viability of light water
  small modular nuclear reactors: General methodology and vendor data.
\newblock \emph{Renewable and Sustainable Energy Reviews} 103: 248--258.
\newblock \doi{10.1016/j.rser.2018.12.041}.

\bibitem[{{Center for Sustainable Systems}(2022)}]{nuclear_facts}
{Center for Sustainable Systems} (2022) Nuclear energy factsheet.
\newblock Technical Report Pub. No. CSS11-15., University of Michigan.

\bibitem[{Ericson(2004)}]{ericson_bvh_2004}
Ericson C (2004) \emph{Real-Time Collision Detection}.
\newblock The {M}organ {K}aufmann Series in Interactive {3D} Technology. {S}an
  {F}rancisco, {CA}, {USA}: {M}organ {K}aufmann {P}ublishers {I}nc.

\bibitem[{Habush and Harris(1968)}]{habush_fsv_1968}
Habush A and Harris A (1968) 330-{MW}(e) {F}ort {S}t. {V}rain high-temperature
  gas-cooled reactor.
\newblock \emph{Nuclear Engineering and Design} 7(4): 312--321.
\newblock \doi{10.1016/0029-5493(68)90064-2}.

\bibitem[{Hamilton and Evans(2019)}]{hamilton_gpu_2019}
Hamilton SP and Evans TM (2019) Continuous-energy {M}onte {C}arlo neutron
  transport on {GPU}s in the {S}hift code.
\newblock \emph{Annals of Nuclear Energy} 128: 236--247.

\bibitem[{Harper et~al.(2020)Harper, Romano, Forget and
  Smith}]{harper_neighbor_2020}
Harper SM, Romano PK, Forget B and Smith KS (2020) Threadsafe dynamic neighbor
  lists for {M}onte {C}arlo ray tracing.
\newblock \emph{Nuclear Science and Engineering} 194(11): 1009--1015.
\newblock \doi{10.1080/00295639.2020.1719765}.

\bibitem[{Hejzlar et~al.(2013)Hejzlar, Petroski, Cheatham, Touran, Cohen,
  Truong, Latta, Werner, Burke, Tandy, Garrett, Johnson, Ellis, McWhirter,
  Odedra, Schweiger, Adkisson and Gilleland}]{hejzlar_terrapower_2013}
Hejzlar P, Petroski R, Cheatham J, Touran N, Cohen M, Truong B, Latta R, Werner
  M, Burke T, Tandy J, Garrett M, Johnson B, Ellis T, McWhirter J, Odedra A,
  Schweiger P, Adkisson D and Gilleland J (2013) Terrapower, {LLC} traveling
  wave reactor development program overview.
\newblock \emph{Nuclear Engineering and Technology} 45(6): 731--744.
\newblock \doi{10.5516/NET.02.2013.520}.

\bibitem[{Johnson et~al.(2021)Johnson, Tognini, Canal, Evans, Jun, Lima, Lund
  and Pascuzzi}]{johnson_celeritas_2021}
Johnson S, Tognini S, Canal P, Evans T, Jun S, Lima G, Lund A and Pascuzzi V
  (2021) Novel features and {GPU} performance analysis for {EM} particle
  transport in the {C}eleritas code.
\newblock \emph{EPJ Web of Conferences} 251(03030).
\newblock \doi{10.1051/epjconf/202125103030}.

\bibitem[{Johnson et~al.(2023)Johnson, Lefebvre and
  Bekar}]{johnson_orange_2023}
Johnson SR, Lefebvre R and Bekar K (2023) {ORANGE}: {O}ak {R}idge {A}dvanced
  {N}ested {G}eometry {E}ngine.
\newblock Technical Report ORNL/TM-2023/3190, Oak Ridge National Laboratory.

\bibitem[{Juarez et~al.(2021)Juarez, Pedroche, Loughlin, Pampin, Martinez,
  De~Pietri, Alguacil, Ogando, Sauvan, Lopez-Revelles, Kol{\v s}ek,
  Polunovskiy, Fabbri and Sanz}]{juarez_iter_2021}
Juarez R, Pedroche G, Loughlin MJ, Pampin R, Martinez P, De~Pietri M, Alguacil
  J, Ogando F, Sauvan P, Lopez-Revelles AJ, Kol{\v s}ek A, Polunovskiy E,
  Fabbri M and Sanz J (2021) A full and heterogeneous model of the {ITER}
  tokamak for comprehensive nuclear analyses.
\newblock \emph{Nature Energy} 6(2): 150--157.
\newblock \doi{10.1038/s41560-020-00753-x}.

\bibitem[{Kos et~al.(2023)Kos, Radulescu, Grove, Villari and
  Batistoni}]{kos_fusion_2023}
Kos B, Radulescu G, Grove R, Villari R and Batistoni P (2023) Comprehensive
  analysis of streaming and shutdown dose rate experiments at {JET} with {ORNL}
  fusion neutronics workflows.
\newblock \emph{Fusion Science and Technology} 79(3): 284--304.
\newblock \doi{10.1080/15361055.2022.2129182}.

\bibitem[{Lee et~al.(2020)Lee, Jung, Zhong, Ortensi, Laboure, Wang and
  DeHart}]{lee_empire_2020}
Lee C, Jung YS, Zhong Z, Ortensi J, Laboure V, Wang Y and DeHart M (2020)
  Assessment of the {G}riffin reactor multiphysics application using the
  {E}mpire micro reactor design concept.
\newblock Technical Report {ANL/NSE-20/23} and {INL/LTD-20-59263}, Argonne
  National Laboratory and Idaho National Laboratory.

\bibitem[{Lieberoth(1968)}]{lieberoth_power_1968}
Lieberoth J (1968) {M}onte {C}arlo technique to solve the static eigenvalue
  problem of the {B}oltzmann transport equation.
\newblock Technical report, INTERATOM, {B}ensberg, {G}ermany.

\bibitem[{Matthews et~al.(2021)Matthews, Laboure, DeHart, Hansel, Andrs, Wang,
  Ortensi and Martineau}]{matthews_empire_2021}
Matthews C, Laboure V, DeHart M, Hansel J, Andrs D, Wang Y, Ortensi J and
  Martineau RC (2021) Coupled multiphysics simulations of heat pipe
  microreactors using {D}ire{W}olf.
\newblock \emph{Nuclear Technology} 207(7): 1142--1162.
\newblock \doi{10.1080/00295450.2021.1906474}.

\bibitem[{Merzari et~al.(2023)Merzari, Hamilton, Evans, Min, Fischer,
  Kerkemeier, Fang, Romano, Lan, Phillips, Biondo, Royston, Warburton, Chalmers
  and Rathnayake}]{merzari_gb_2023}
Merzari E, Hamilton S, Evans T, Min M, Fischer P, Kerkemeier S, Fang J, Romano
  P, Lan YH, Phillips M, Biondo E, Royston K, Warburton T, Chalmers N and
  Rathnayake T (2023) Exascale multiphysics nuclear reactor simulations for
  advanced designs.
\newblock In: \emph{Proceedings of the International Conference for High
  Performance Computing, Networking, Storage and Analysis}, SC '23. New York,
  NY, USA: Association for Computing Machinery.
\newblock \doi{10.1145/3581784.3627038}.

\bibitem[{Mulder and Boyes(2020)}]{mulder_x-energy_2020}
Mulder E and Boyes W (2020) Neutronics characteristics of a 165 {MW}th {X}e-100
  reactor.
\newblock \emph{Nuclear Engineering and Design} 357.
\newblock \doi{10.1016/j.nucengdes.2019.110415}.

\bibitem[{Nelson et~al.(2022)Nelson, Smith, Karriem, Navarro, Denbrock, Wahlen
  and Grimm}]{nelson_niowave_2022}
Nelson NB, Smith MBR, Karriem Z, Navarro J, Denbrock CP, Wahlen RN and Grimm TL
  (2022) Radiation shielding analysis of {N}iowave’s uranium target assembly
  2 ({UTA}-2) facility for molybdenum-99 production.
\newblock Technical Report ORNL/TM-2021/2269, Oak Ridge National Laboratory.
\newblock \doi{10.2172/1878714}.

\bibitem[{Nu{S}cale~Power(2023)}]{nuscale}
Nu{S}cale~Power L (2023) About us.
\newblock \urlprefix\url{https://www.nuscalepower.com/en/about}.

\bibitem[{{O}ak {R}idge {L}eadership~{C}omputing
  {F}acility(2023{\natexlab{a}})}]{frontier}
{O}ak {R}idge {L}eadership~{C}omputing {F}acility (2023{\natexlab{a}})
  Frontier.
\newblock
  \urlprefix\url{https://www.olcf.ornl.gov/olcf-resources/compute-systems/frontier}.

\bibitem[{{O}ak {R}idge {L}eadership~{C}omputing
  {F}acility(2023{\natexlab{b}})}]{summit}
{O}ak {R}idge {L}eadership~{C}omputing {F}acility (2023{\natexlab{b}}) Summit:
  {O}ak {R}idge {N}ational {L}aboratory's 200 petaflop supercomputer.
\newblock
  \urlprefix\url{https://www.olcf.ornl.gov/olcf-resources/compute-systems/summit}.

\bibitem[{OpenClipart(2023)}]{map}
OpenClipart (2023) Outline map of {A}merican states.
\newblock \urlprefix\url{https://freesvg.org/outline-map-of-american-states}.

\bibitem[{Pandya et~al.(2016)Pandya, Johnson, Evans, Davidson, Hamilton and
  Godfrey}]{pandya_shift_2016}
Pandya TM, Johnson SR, Evans TM, Davidson GG, Hamilton SP and Godfrey AT (2016)
  Implementation, capabilities, and benchmarking of {S}hift, a massively
  parallel {M}onte {C}arlo radiation transport code.
\newblock \emph{Journal of Computational Physics} 308: 239--272.
\newblock \doi{10.1016/j.jcp.2015.12.037}.

\bibitem[{Radel and Van~Abel(2016)}]{radel_shine_2016}
Radel T and Van~Abel E (2016) Validation of {MCNP5} for use in calculating
  temperature coefficients of reactivity for the {SHINE} system.
\newblock \emph{Transactions of the American Nuclear Society} 114(1).

\bibitem[{Smith(2017)}]{smith_smr_2017}
Smith K (2017) {NuScale} small modular reactor ({SMR}) progression problems for
  the {ExaSMR} project.
\newblock Milestone Report WBS 1.2.1.08 ECP-SE-08-43, Exascale Computing
  Project.

\bibitem[{von Mises and Pollaczek-Geiringer(1929)}]{mises_power_1929}
von Mises R and Pollaczek-Geiringer H (1929) Praktische verfahren der
  gleichungsaufl{\"o}sung.
\newblock \emph{Zamm-zeitschrift Fur Angewandte Mathematik Und Mechanik} 9:
  152--164.

\bibitem[{W{\"a}chter and Keller(2006)}]{wachter_bih_2006}
W{\"a}chter C and Keller A (2006) Instant ray tracing: The bounding interval
  hierarchy.
\newblock In: Akenine-Moeller T and Heidrich W (eds.) \emph{Symposium on
  Rendering}. The Eurographics Association.
\newblock ISBN 3-905673-35-5.
\newblock \doi{10.2312/EGWR/EGSR06/139-149}.

\bibitem[{Wald(2007)}]{wald_sah_2007}
Wald I (2007) On fast construction of {SAH}-based bounding volume hierarchies.
\newblock In: \emph{2007 {IEEE} Symposium on Interactive Ray Tracing}. pp.
  33--40.
\newblock \doi{10.1109/RT.2007.4342588}.

\bibitem[{West et~al.(1979)West, Petrie and Fraley}]{west_keno_1979}
West JT III, Petrie LM and Fraley SK (1979) {KENO-IV/CG}, the combinatorial
  geometry version of the {K}eno {M}onte {Carlo} criticality safety program.
\newblock Technical Report {NUREG/CR-0709, ORNL/NUREG/CSD-7}, Oak Ridge
  National Laboratory.

\bibitem[{Zhang et~al.(2021)Zhang, Alawneh and Rogers}]{zhang_virtual_2023}
Zhang M, Alawneh A and Rogers TG (2021) Characterizing massively parallel
  polymorphism.
\newblock In: \emph{2021 {IEEE} International Symposium on Performance Analysis
  of Systems and Software ({ISPASS})}. pp. 205--216.
\newblock \doi{10.1109/ISPASS51385.2021.00037}.

\end{thebibliography}
\end{document}